\documentstyle[aps,pre,epsfig,graphicx]{revtex}
\newcommand{\be}{\begin{equation}}
\newcommand{\bea}{\begin{eqnarray} \nonumber}
\newcommand{\ee}{\end{equation}}
\newcommand{\eea}{\end{eqnarray}}

 \def\(({\left(}
 \def\)){\right)}
\def\[[{\left[}
\def\]]{\right]}
\def\bi{\bibitem}
\def \form#1 {eq. (\ref{#1}) }
\def \parziale#1#2  {{\partial {#1} \over \partial {#2}}}
\def \atanh{\hbox{atanh}}
\def \cM{{\cal M}}
\def \cN{{\cal N}}
\def \cC{{\cal C}}
\def\la{\langle}
\def\ra{\rangle}
\def \Tr {\mbox{Tr}}
\def \ba#1 {\overline{#1}}
\def  \bh {{\bf h}}
\def  \bg {{\bf g}}

\def  \s {\sigma}
\begin{document}

\title{ The Bethe lattice spin glass revisited}

\author{Marc M\'ezard}
\address{Laboratoire de Physique Th\'eorique et Mod\`eles Statistiques
\\
Universit\'e Paris Sud, Bat. 100, 91405 Orsay {\sc cedex}, France \\
mezard@ipno.in2p3.fr}
\author{Giorgio Parisi}
\address{Dipartimento di Fisica, Sezione INFN and Unit\`a INFM ,\\
Universit\`a di Roma ``La Sapienza'',\\
Piazzale Aldo Moro 2,
I-00185 Rome (Italy)\\
giorgio.parisi@roma1.infn.it}

\date{\today}

\maketitle

\begin{abstract}
So far the problem of a spin glass on a Bethe lattice has been solved
only at the replica symmetric level, which is wrong in the spin glass
phase. Because of some technical difficulties, attempts at deriving 
a replica symmetry breaking solution have been confined to some 
perturbative regimes, high connectivity lattices or temperature close to the
critical temperature.
Using the cavity method, we propose a general non perturbative
 solution of the Bethe lattice spin glass problem at a level of approximation
which is equivalent to a one step replica symmetry breaking solution. 
The results compare well with numerical simulations.
The method can be used for many finite connectivity problems appearing
in combinatorial optimization.
\end{abstract}

\section{Introduction}
The spin glass problem has been around for twenty five years, but its understanding has turned
out to be remarkably complicated.
It is generally considered as solved 
 only in its fully connected version
introduced by Sherrington and Kirkpatrick \cite{SK}. The first
consistent solution  was derived 
with the replica method \cite{parsol,MPV} and it was then confirmed
 using a probabilistic
approach, the cavity method, which  avoids the strange (and powerful)
mathematical subtleties of the replica approach \cite{cavity,MPV}. A rigorous
proof of the validity of the solution is still lacking, in spite of
recent progress\cite{guerra,aizenman,talagrand}.

A slightly more realistic theory of spin glasses, still of the mean field type,
  deals with the situation in which each spin interacts  only with 
a finite number of neighbours. 
Models of this type include the spin glass on a Cayley tree,
 a Bethe lattice 
and a disordered random lattice with fixed or with 
fluctuating connectivity.
There are many motivations for studying such problems. On
 one hand one may hope to get a better knowledge on the finite dimensional
problem, since these models include a notion of neighborhood
which is absent in the infinite range case.
 But another motivation is the possibility 
to  solve these problems using different methods, like 
iterative methods which are typical of statistical mechanics on 
tree-like structures. In fact the cavity method is a generalization
of the Bethe Peierls iterative method to the case in which 
there may exist several pure states, and it is therefore very natural to work 
out the details of this generalisation, and to test its validity.
 Another important aspect comes from the connection between
the statistical mechanics of disordered systems and the optimization problems:
many of the interesting random optimization problems turn out to
have a finite connectivity structure. This is the case for instance
of the travelling salesman problem \cite{tsp}, the matching \cite{matching}
the graph partitioning \cite{fuand,banshesou} 
or the K-satisfiability problem \cite{Ksat}.

While the problems of a spin-glass on  tree-like lattices
were naturally  studied very soon after the discovery of spin glasses,
the present status of the knowledge on these systems  is still rather poor
compared to that on the SK model. A lot of efforts have been devoted to the
simple Bethe Peierls method which builds up a solution 
in terms of the distribution of local
magnetic fields \cite{klein,katsura,nakanishi,bowman,thouless,chayes}. 
However this simple iterative solution, 
which may be relevant for a Cayley tree
with a certain type of boundary conditions \cite{chayes}
 is wrong for  the Bethe lattice spin-glass. When the
replica formalism is used, this simple iterative solution turns out to be
 equivalent to the replica symmetric (RS) solution. However
one knows that there exists a  replica symmetry breaking (RSB)
instability similar to the one found in the SK model 
\cite{thouless,motti,motti2,dewar,Lai}.
 Unfortunately, in the replica formalism, the RSB
solution could be found only in some rather limited regimes: expansion around the
high connectivity (SK-like) limit \cite{DDGold}, or close to 
the critical temperature
\cite{motti2}. The main problem encountered in all these attempts
 is a very general one,
common to all disordered problems with a finite connectivity. Roughly
speaking it can be summarized as follows: the distribution of
local fields, even within one pure state,
 is not a simple Gaussian as in the infinite
range problems, but a more complicated function \cite{matching,orland,mp86}. 
When one takes
into account the existence of several pure states, the natural order parameter,
even within  the simplest ``one step'' replica symmetry breaking solution,
 becomes the probability distribution of these local field probabilities
\cite{monasson98}. This
 is a functional order parameter which  is difficult to handle. 
Several interesting attempts at solving the one step RSB equations have been
done in the past \cite{WonShe,goldschmidt_gpref,BiMoWe},
but they all 
 restricted this functional order parameter to
some particular subspace, within which a variational aproach was used.

In this paper we present an improved 
solution of the Bethe lattice spin glass.
This solution is nothing but the application to this problem of
 the cavity method, treated at a level which is equivalent to the  one step RSB
solution. It
is valid for any connectivity and any
temperature. 
In the next section we discuss the various tree-like lattices which are usually
studied and precise  our definition of the Bethe lattice problem. 
In sect. \ref{sect_rs} we recall the basic steps of the
simple Bethe-Peierls approach, and we discuss its instability 
in sect. \ref{sect_instab}. 
In sect. \ref{sect_1rsb}
where  we discuss the formalism of the   cavity approach at the one step RSB level.
Sect. \ref{sect_algo} describes the algorithm used to determine the
distribution of local fields within this approach. 
The implementation of the algorithm is discussed in sect. \ref{numerics},
where we derive explicit results for a lattice with six neighbours per point and 
compare the analytic prediction to those of numerical simulations.
Finally, sect. \ref{conclusion} contains a brief discussion and mentions the 
perspectives.

\section{The Bethe lattice}
We consider a system of $N$ Ising spins, $\sigma_i= \pm1$, $i \in \{ 1,...,N\}$,
interacting with random couplings, the energy being:
\be
E=-\sum_{<ij>} J_{ij} \sigma_i \sigma_j \ .
\ee
The sum is over all links of a lattice.  For each link $<ij>$ the coupling $J_{ij}$ is an 
independent random variable chosen with the same probability distribution $P(J)$.  The 
various types of tree-like lattices which have been considered are:
\begin{itemize}
\item A) The Cayley tree: starting from a central site $i=0$, one builds a first 
shell of $k+1$ neighbours. Then each of the first shell spins is connected
to $k$ new neighbours in the second shell etc... until one reaches the $L$'th
shell which is the boundary.
 There is no overlap among the new neighbours, so that the graph is a tree.
\item B)  The random graph with fluctuating connectivity: for each pair of
indices $(ij)$, a link is present with probability $c/N$ and absent with probability
$1-c/N$. The number of links connected to a point is a random variable with 
a Poisson distribution, its mean being equal to $c$.
\item C)  The random graph with fixed connectivity, equal to $k+1$. 
The space of allowed
graphs are all graphs such that the number of links connected 
to each point is equal to 
$k+1$. The simplest choice, which we adopt here,
 is the case where  every such graph has the same probability.
\end{itemize}  

On a Cayley tree a finite fraction of the 
total number of spins lie on the boundary.  The 
Cayley tree is thus a strongly inhomogeneous system, 
the properties of which are often 
remote from those of a usual finite dimensional problem.  
For this reason people generally 
consider instead a {\it Bethe lattice}, which consists 
of a subset of the Cayley tree 
containing the first $L'$ shells.  Taking the limits 
$L \to \infty$, $L' \to \infty$ with 
$L/L' \to \infty$ allows to isolate the central part of 
the tree, away from the boundary.  
This procedure is OK when one considers a ferromagnetic 
problem.  In the case of a spin 
glass this definition of the Bethe lattice is not free 
from ambiguities: one can not 
totally forget the boundary conditions which are imposed 
on the boundary of the Cayley 
tree, since they are fixing the degree of frustration\cite{Lai}.  
For this reason we prefer to 
define the Bethe lattice as the random lattice with fixed 
connectivity (lattice C defined 
above).  Clearly on such a graph the local structure is 
that of a tree with a fixed 
branching ratio.  Small loops are rare, the typical size 
of a loop is of order $\log N$.  
Therefore in the large $N$ limit the random graph with 
fixed connectivity provides a well 
defined realisation of a Bethe lattice, i.e. a statistically 
homogeneous, locally 
tree-like structure.  This is the lattice which we study 
in this paper (the case of 
fluctuating connectivities will be studied in a forthcoming 
work).  Numerical simulations 
of this system can be found in \cite{banshesou,Lai,PARI,MAZU}.

Historically, spin glasses on the Bethe lattice and on 
diluted lattices with a fixed 
finite connectivity (type C) were often discussed as 
a separate issue.  The reason for 
these separate discussions of the same problem is the 
type of techniques which are used.  
Generally speaking the Bethe lattice papers rely on the 
Bethe Peierls method while the 
random lattice papers use the replica method.  One 
exception is the use of the cavity 
method for the random lattice case \cite{mp86,goldschmidt_gpref}.
Hereafter we shall basically develop the iterative/cavity approach, but we shall
also mention at each step its connexions to the replica approach.

\section{The simple Bethe-Peierls 'solution'}
\label{sect_rs}
This section will give a brief review to the  standard approach to the
spin glass on the Bethe lattice, defined as lattice C in the above 
classification. This 
solution is wrong, because, as we shall see later, it does not consider 
the phenomenon of 
replica symmetry breaking, however it sets the stage for 
the correct solution that will be 
presented in the next section.

\subsection{The iterative approach}\label{sect_rs_iter}

\begin{figure}
\hbox{\epsfig{figure=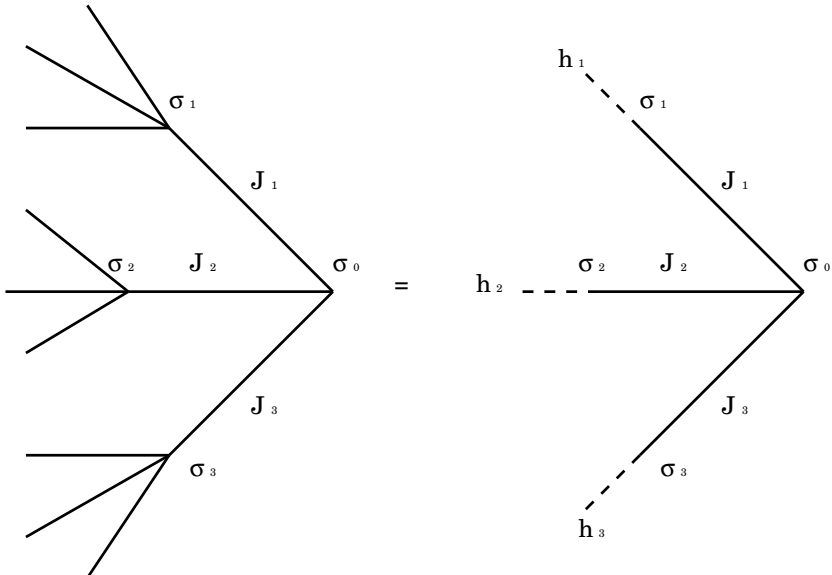,width=11cm
}}
\caption{ The merging of $k$ branches off the tree (here $k=3$) onto
the spin $\s_0$. The cavity field  $h_i$ is the total field acting on spin $\s_i$
in the absence of the central spin $\s_0$.}
\label{fig_branch}
\end{figure}

As is well known, on  tree-like structures the problem can be 
solved by iteration.
Let us consider in general the merging of $k$ branches  of a  tree 
onto one  site $\s_0$ as in fig.\ref{fig_branch}. The partition function 
can be computed exactly if
one introduces, for each of the outside spins $\s_i, i\in\{1 ,...,k\} $, 
the effective
"cavity" field $h_i$ representing the action onto the spin $\s_i$ of all the
other spins, {in the absence of the central spin} $ \s_0$. 
In other words the magnetization
of the $i$-th spin in absence of the central spin is given 
by $m_{i}=\tanh(\beta h_{i})$. The variables 
$h_{i}$ (and consequently the variables $m_{i}$) are 
uncorrelated in the limit $N \to 
\infty$. It is crucial to consider the magnetization 
before the introduction of the spin $ 
\s_0$, because after its introduction  
all the spins that are 
coupled to the spin $ \s_0$ become correlated. 

Calling 
$J_i$ the coupling between spins $\s_0$ and $\s_i$, the partition 
function of the spin $ \s_0$
is expressed as
\be
 \sum_{\s_0,\s_1,...\s_k} \exp\((\beta \s_0 \sum_{i=1}^k J_i \s_i +\beta 
\sum_{i=1}^k h_i \s_i\)) \ .
\label{simple_iter}
\ee
Let us recall here the basic identity which allows 
to forward the effect of the fields $h_i$ onto spin $\s_0$, and which
 is used repeatedly in this work. For an Ising spin $\s_0=\pm 1$, one has:
\be
\sum_{\s=\pm 1} \exp\((\beta \s_0 J \s + \beta h \s\))= 
c(J,h) \exp(\beta u(J,h) \s_0)
\label{propag_h}
\ee
where we define the two functions $u$ and $c$ as:
\be
u(J,h)= {1 \over \beta } \atanh \[[ \tanh(\beta J) \tanh(\beta h)\]]
 \ \ \ ; \ \ \ c(J,h)= 2 {\cosh (\beta J) \cosh (\beta h) \over
 \cosh (\beta u(J,h))}
\ee
The magnetization on site $0$ is thus
$m_0=<\s_0>=\tanh(\beta h_0)$,
where
\be
h_0=  \sum_{i=1}^k u(J_i,h_i) \ \ \ ; \ \ \ 
\label{iterh}
\ee
 From this equation one gets
the basic recursion relation for the probability density 
$Q(h)$ 
of local fields:
\be
Q(h)= E_J\int
\prod_{i=1}^{k} \[[dh_i Q(h_i)\]] \ 
 \delta\((h- \sum_{i=1}^k u( J_i,h_i)\))  \ .
\label{RICOR}
\ee
Here and throughout the paper, we denote by $E_J$ the expectation value
with respect to all the exchange coupling constants $J_i$:
$E_J = \int \prod_i \[[dh_i  P(J_i)\]]$ .

It will be useful for future use to introduce the
probability  distribution
$R(u)$ of the propagated field variable $u(J,h)$:
\be
R(u)=  E_J \int Q(h) dh \ \delta ( u-u(J,h))
\label{Rdef}
\ee
The field distribution is nothing but the 
convolution: $ Q(h)= \int du_1\ldots du_k \ R(u_1) \ldots R(u_k) \delta (u_1+\ldots
+u_k -h)$.
 
In order to relate the true distribution 
of local fields, $Q_{t}(H)$,   to  the distribution $Q(h)$
of local fields on one branch\footnote{We  denote by upper 
case letters the true local fields and by  small case letters 
the local fields on one 
branch.},
 one needs to consider the merging of $k+1$ branches onto 
one site. The true local field $H_0$  on a given site $0$ 
is simply given by a sum of contributions from each of its $k+1$ 
neighbours,
\be
H_{0}=  \sum_{j=1}^{k+1} u(J_j,h_j) \ ,
\label{iterhkp1}
\ee
where as before $h_j$ is the local field on $j$ in the absence of the 
spin $s_0$.  This 
gives the distribution of true local fields $Q_{t}(H)$ as the 
convolution: 
\be
Q_t(H)=  \int \prod_{i=1}^{k+1} \[[du_i
R(u_i) \]] \delta\((H- \[[\sum_{i=1}^{k+1} u_i\]] \)) \ .
\label{Qt}
\ee

Let us now  compute the internal energy with this method.  We
 add a new link \cite{substract} with a coupling constant $J_{ij}$
between two spins $\s_i$ and $\s_j$, where the local 
fields in the absence of the new link
are respectively $h_i^{(j)}$ and $h_j^{(i)}$.
Then the energy of this link is:
\be
E_{ij}=- J_{ij}\langle\s_i \s_j \rangle\ , 
\ee
where the expectation value is computed using the Hamiltonian 
$H_{ij}(\s_i,\s_j)$, which is given by
\be
H_{ij} (\s_i,\s_j)=-\(( J_{ij} \s_i \s_j +h_i^{(j)}\s_i+h_j^{(i)}\s_j\)) \ .
\ee
A simple computation shows that
\be
E_{ij}=-J_{ij} {\tanh( \beta J_{ij}) + 
\tanh( \beta h_i^{(j)} ) \tanh( \beta h_j^{(i)} ) 
\over
1+ \tanh (\beta J_{ij})\tanh( \beta h_i^{(j)})  \tanh (\beta h_j^{(i)} )} \ .
\label{ene}
\ee

Computing the total free energy of the system is slightly more involved.
 Using the fact that the Bethe-Peierls
approximation is exact on the tree-like lattices one can write the free
energy as the sum of site and bond contributions 
\cite{katsura,nakanishi,bowman}:
\be
F=-k \sum_i F_i^{(1)}+
 \sum_{<ij>} F_{<ij>}^{(2)} \ ,
\label{freetot}
\ee
where the contribution from the bond $ij$
is
\be
-\beta F_{<ij>}^{(2)}= \ln \sum_{\s_i,\s_j} \exp \((
-\beta H_{ij} (\s_i,\s_j) \))
\label{free2}
\ee
and that from the site $i$ is:
\be
-\beta F_{i}^{(1)}=  \ln \sum_{\s_i} \exp\((\beta H_i \s_i\)) \ ,
\label{free1}
\ee
where $H_i$ is the total spin acting on spin $\s_i$. One can prove the
validity of the expression (\ref{freetot}) by the following
two steps: 1) 
it clearly gives the correct free energy
at high temperature; 2) using the fact that $\sum_{j(i)}h_i^{(j)}=k H_i$,
where the sum is over all the neighbours $j$ of site $i$,
one  finds that $ \partial (\beta F)/\partial \beta$
gives back the correct expression for the internal energy obtained in (\ref{ene}).
We notice {\sl en passant}   that this free energy is nothing but
the generalization to a finite coordination number of the TAP free 
energy (and reduces to the usual TAP free energy in the limit of infinite 
coordination number)\cite{nakanishi,bowman}. 

The Edwards-Anderson order parameter \cite{EA}, 
$q =(1/N) \sum_i <\s_i>^2$  can 
be written as the magnetization 
squared of a spin coupled to $k+1$ neighbours and is given by
\be
q= \int d H Q_t(H) \  [\tanh^2(\beta H)] \ .
\ee
We shall also compute the link overlap, $q^{(l)}=(2/(N(k+1)) \sum_{<ij>}
 <\s_i \s_j>^2$, which is deduced from the $Q(h)$ distribution as:
\be
q^{(l)}= E_J \int dh dh' Q(h) Q(h')
\(({\tanh( \beta J) + 
\tanh( \beta h ) \tanh( \beta h' ) 
\over
1+ \tanh (\beta J)\tanh( \beta h)  \tanh (\beta h' )}\))^2
\ee

\subsection{A variational formulation}
We have just seen in the previous section that, if one  neglects 
the possibility of RSB, all the thermodynamic quantities of
 the Bethe lattice spin glass can be computed in terms of the probability 
distribution 
$Q(h)$ of the effective field $h$. This probability distribution is obtained
by solving the self-consistency equation (\ref{RICOR}).

\begin{figure}
 \includegraphics[width=0.37\textwidth]{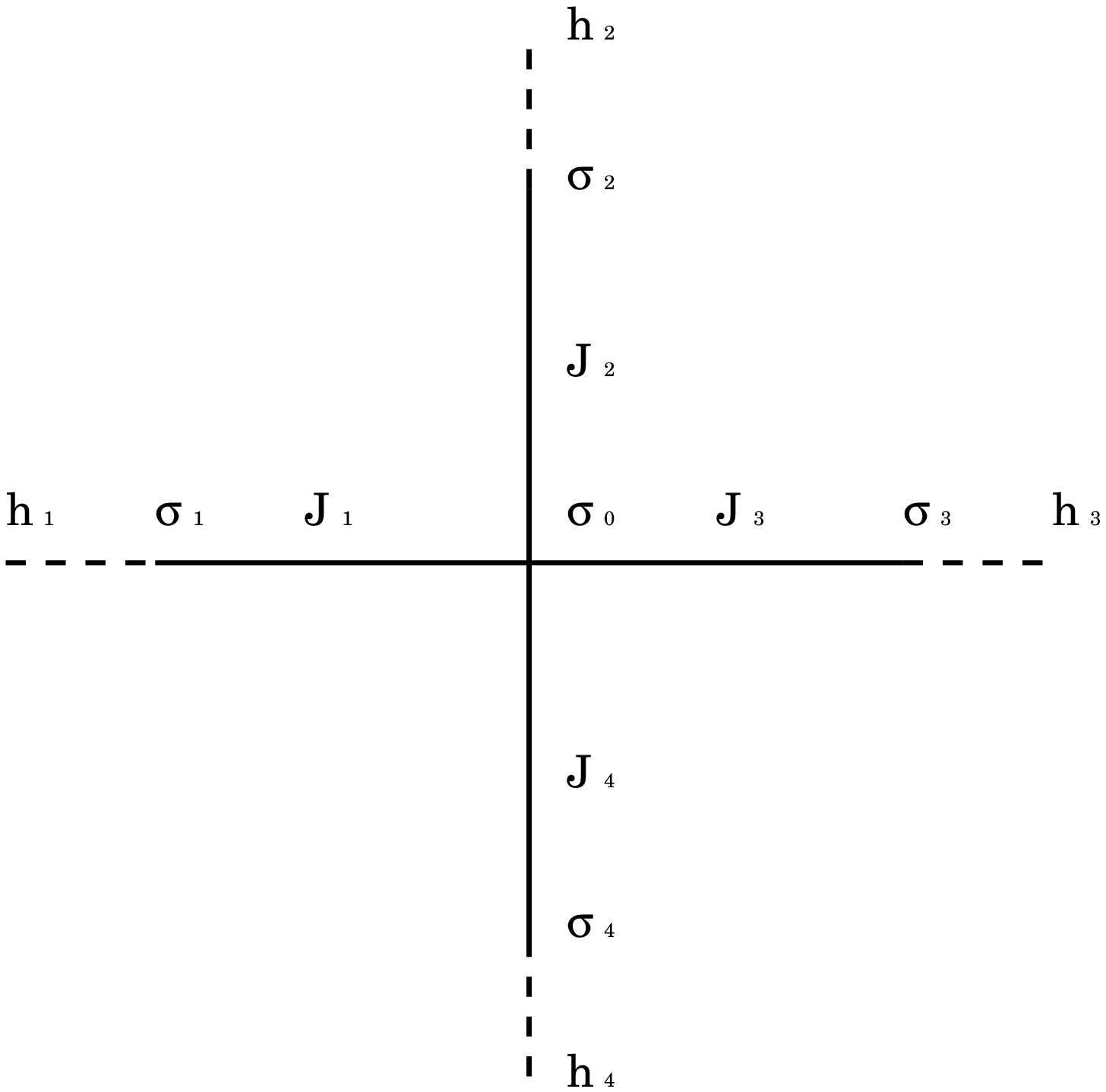}
 \includegraphics[width=0.57\textwidth]{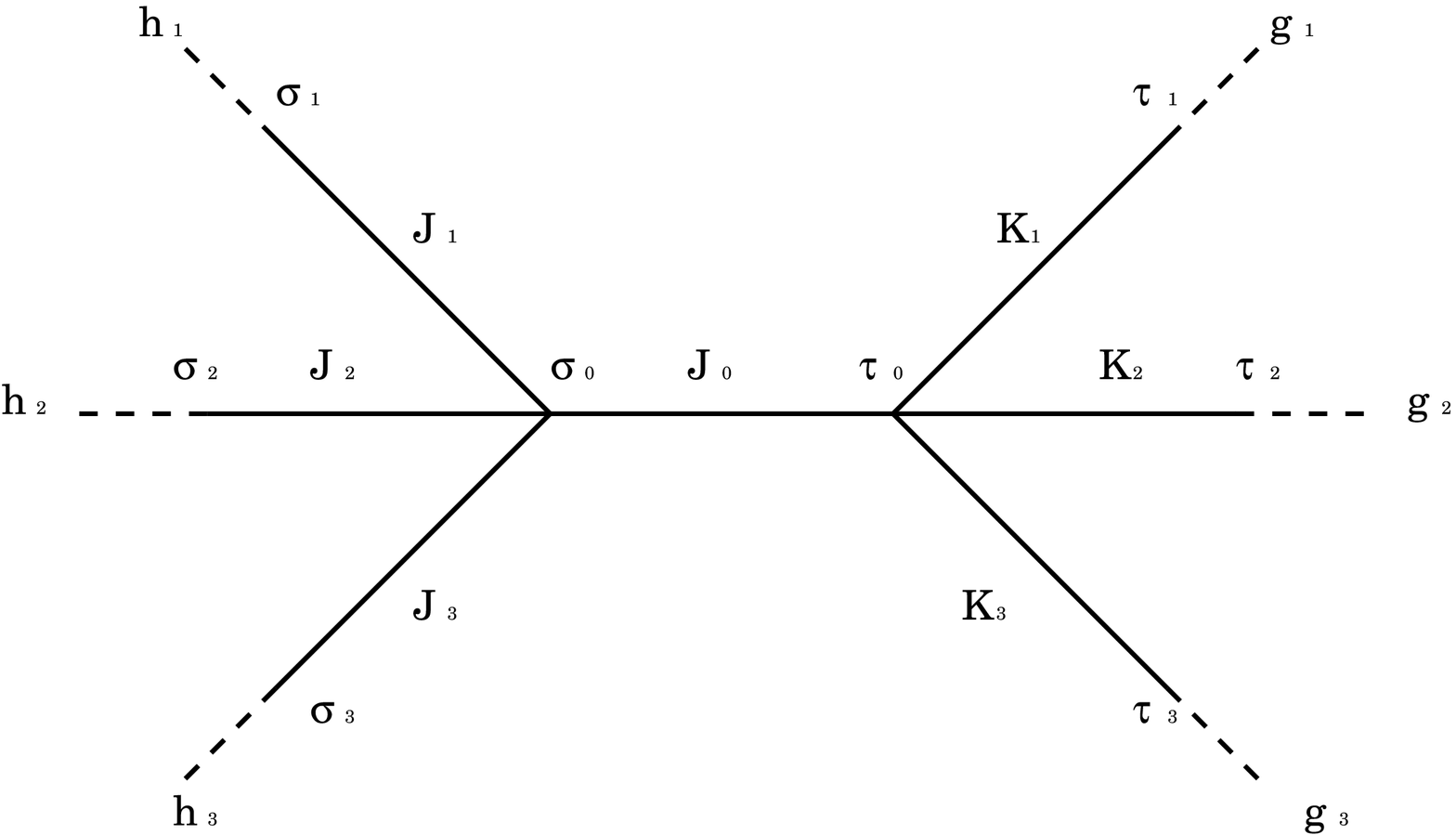}

\caption{ A pictorial representation of the two contributions (\ref{bigfree_det})
 to the free energy. The `site' contribution on the top is obtained
by merging $k+1$ lines onto one site (here $k=3$), and the `bond' contribution
pictured on the bottom figure is obtained by adding one new link $J_0$, and two
new spins $\s_0$ and $\tau_0$, to the lattice, together with the other 
$k$ branches arriving onto each of these spins. }
\label{fig_free}
\end{figure}

It is interesting for many reasons, some of which will become clear later, 
to reformulate this problem in a variational way. One can write a free energy 
 $F[Q]$, which is a functional of the probability 
distribution $Q(h)$, such that: 
\begin{enumerate}
\item The equation
$
{\delta F}/{\delta Q(h)} =0
$
is equivalent to the self-consistency  equation (\ref{RICOR}) for $Q(h)$.

\item Calling $Q^{*}$  the solution of the previous equation, 
the  equilibrium free energy (\ref{freetot}) is equal to $F[Q^{*}]$.
\end{enumerate}
This free energy functional is given by:
\bea
{F[Q] \over N}=&&\frac{k+1}{2}  \int \prod_{i=1}^k
\[[ dh_i dg_i Q(h_{i})Q(g_{i}) \]] \ 
F^{(2)}(h_1 \ldots h_k,g_1\ldots g_k ) \\
&-& k \int\prod_{i=1}^{k+1}\[[ dh_i  Q(h_{i}) \]] \ 
 F^{(1)}(h_1\ldots h_{k+1} ) \,
\label{bigfree} 
\eea
where
\bea
-\beta F^{(1)}(h_1\ldots h_{k+1})&=&E_J 
\ln\(( \[[\prod_{i=1}^{k+1} {1 \over d(J_i,h_i)} \]]
\sum_{\s_0,\s_1,\ldots \s_{k+1}}
\exp\[[\beta \s_0 \sum_{i=1}^{k+1}J_i \s_i+\beta \sum_{i=1}^{k+1}h_i \s_i\]]
 \)) \ , \\ \nonumber
-\beta F^{(2)}(h_1 \ldots h_k,g_1\ldots g_k)&=&E_J E_K
\ln\(( \[[\prod_{i=1}^{k} {1 \over d(J_i,h_i) d(K_i,g_i)} \]]
\sum_{\s_0,\s_1,\ldots \s_{k}} \  \sum_{\tau_0,\tau_1,\ldots \tau_{k}} \right.
\\
&&
\left.
\exp\[[\beta J_0 \s_0 \tau_0 +\beta\s_0 \sum_i J_i \s_i
+ \beta \sum_i h_i \s_i +\beta\tau_0\sum_i K_i \tau_i+\beta\sum_i g_i \tau_i \]]
\))
\label{bigfree_det}
\eea
These two expressions are represented pictorially in fig. \ref{fig_free}.
In this formula the function $d(J,h)$ is an arbitrary positive function,
since its contributions to the two pieces 
 $F^{(1)}$ and $F^{(2)}$ cancel. We shall mainly use it with
$d(J,h)=c(J,h)$, which allows an easy  connection
with the expression (\ref{freetot}), but some other choice will also be useful in 
order to make contact with the result of the  
replica method, as  we shall see below.

Let us now show that this free energy has the desired properties.
In order to check that $Q^*(h)$ is a stationarity point of the free energy
in the space of normalized probability distributions $Q(h)$(
such that $\int dh Q(h)=1$),
we need to show that
    $\delta F/\delta Q(h)=constant$ when $Q=Q^*$.
This functional derivative is equal to
\bea
  {1 \over N}&&  {\delta F[Q] \over \delta Q(h)}=
    k(k+1) \int dh_2 \ldots dh_k Q(h_{2}) \ldots Q(h_{k})\\ 
&&\((\int dg_1\ldots dg_k Q(g_{1})\ldots  Q(g_{k}) \  
    F^{(2)}(h_1 \ldots h_k,g_1\ldots g_k)
    -\int dh_{k+1} Q(h_{k+1})F^{(1)}(h,h_2\ldots h_{k+1})\))
\ .
\label{funcder}
\eea
Using (\ref{propag_h}), one easily sees that, 
if $Q(h)=Q^*(h)$ satisfies the self consistency equation 
(\ref{RICOR}), one has, for any $h_1 \ldots h_k$:
\be
\int dg_1\ldots dg_k Q^*(g_{1})\ldots Q^*(g_{k}) 
 F^{(2)}(h_1 \ldots h_k,g_1\ldots g_k)
=
A+ \int dh_{k+1} Q^*(h_{k+1})  F^{(1)}(h_1\ldots h_{k+1})  \ ,
\ee
where $A$ is a constant (independent of $h_1 \ldots h_k$), given by
\be
A =(-1/\beta)\int dg_0 Q^*(g_0) E_J \(( \ln \[[d(J_0,g_0)\]]
+k\ln \[[c(J_0,g_0)/d(J_0,g_0)\]] \)) \ .
\ee
This shows that the functional derivative (\ref{funcder}) is  a constant.
The repeated use of (\ref{propag_h}) allows to show similarly that the
saddle point free energy $F[Q^*(h)]$ is indeed equal to the free energy
(\ref{freetot}) (the factor $(k+1)/2$ in (\ref{bigfree}) is nothing 
but the number of links per site).

As an extra check, one can see that 
 the derivative of $\beta F[Q]$ 
we respect to $\beta$ gives the internal energy of the previous section 
(the derivative is quite simple if we 
absorb most of the $\beta$'s redefining the $h$ and notice that the only explicit 
dependence on $\beta$ comes from the term $\beta J$).

It is interesting to note that, using the basic recursion relation (\ref{propag_h}),
and the special choice $d(J,h)=2 \cosh(\beta h)$, 
we can write the free energy under the simple form:
\be
{F[Q] \over N}=\frac{k+1}{2}  \int \prod_{i=1}^k
\[[ dh_i  Q(h_{i}) \]] \ 
F^{(1')}(h_1 \ldots h_k ) 
-\frac{k-1}{2}  \int\prod_{i=1}^{k+1}\[[ dh_i  Q(h_{i}) \]] \ 
 F^{(1)}(h_1\ldots h_{k+1} ) \,
\label{bigfree_simple} 
\ee
where
\be
-\beta F^{(1')}(h_1\ldots h_{k})=E_J 
\ln\(( \[[\prod_{i=1}^{k} {1 \over 2 \cosh(\beta h_i)} \]]
\sum_{\s_0,\s_1,\ldots \s_{k}}
\exp\[[\beta \s_0 \sum_{i=1}^{k}J_i \s_i+\beta \sum_{i=1}^{k}h_i \s_i\]]
 \))
\ee
One must keep in mind that this expression for the free  
energy is correct only if $Q$ 
satisfies  eq. (\ref{RICOR}) and should {\sl not} be used as a
variational free energy (see also
next section).

\subsection{Equivalence with the replica formalism}
We shall not present here the details of the replica approach to this problem,
for which we refer the reader to \cite{motti,motti2,WonShe,DDGold,monasson98}.
Let us recall the main results of \cite{DDGold}. 
In the replica approach one introduces 
a probability distribution $\rho(\sigma)$, where the variables $\s$ are $n$ Ising 
variables ($n$ eventually goes to zero). One can introduce a free energy functional 
$F_{rep}[\rho(\sigma)]$. The equilibrium free energy is given by
$F(\beta) =F_{rep}[\rho^{*}]$, where $\rho^{*}$ is the solution
 of the stationarity equation
${\delta F_{rep}}/{\delta \rho}=0$.

The expression of the free energy functional in replica space 
can be derived following 
exactly the same steps as in the previous section. 
The final result is \cite{DDGold}
\be
-\beta  n {F_{rep}[\rho] \over N} =
\frac{k+1}{2}\ln\((\Tr_{\sigma,\tau}\[[\rho(\sigma)^{k}\rho(\tau)^{k} 
\exp \((\sum_{a=1}^n\beta J \sigma_{a}\tau_{a}\))\]]\))
 -k 
\ln\((\Tr_{\sigma}\[[\rho(\sigma)^{k+1}\]]\))
\label{repfree}
\ee
where $\Tr_{\sigma}$ denotes the average over the $2^{n}$ configurations of 
the  variables $\sigma$ or $\tau$, and the correct result is obtained in the
$n \to 0$ limit. The same value for the free energy is obtained if we multiply the 
function $\rho$ by a constant, so that the $\rho$ does not need to be normalized, 
although it is more convenient to work with a normalized $\rho$.
The advantage of the replica approach is that the system is homogenous and the 
distribution $\rho(\sigma)$ is  the same in all the points. 
This advantage is partially 
compensated by the fact the the number of variables $n$ is going to zero.

As can be readily checked, $\rho$ satisfies a very simple equation:
\be
\rho(\sigma)=
{ 
E_{J} \Tr_{\tau} \[[\rho(\tau)^{k} 
\exp \((\sum_{a=1}^n\beta J \sigma_{a}\tau_{a}\))\]]
\over
\Tr_{\tau} \[[\rho(\tau)^{k}\]]
}  \ .
\label{rhoeq}
\ee
Using this equation the free energy (\ref{repfree})  can be simplified to
\cite{DDGold}:
\be
-\beta  n {F_{rep}[\rho] \over N} =
\frac{k+1}{2}\ln\((\Tr_{\sigma}\[[\rho(\sigma)^{k}\]] \))
 -\frac{k-1}{2} 
\ln\((\Tr_{\sigma}\[[\rho(\sigma)^{k+1}\]]\))
\label{repfree_simple}
\ee
where as before this new form of the free energy cannot be used in a variational 
formulation.
 The  result (\ref{repfree}) for the replicated free energy is correct in general,
 whether the replica symmetry is broken or not.
The problem is to find the solution $\rho^*(\s)$.
In the replica symmetric situation this task is easy:
 the $\rho(\sigma)$ is a function of only 
$
\Sigma=\sum_{a}\sigma_{a}
$.
We can thus write in general:
\be
\rho(\sigma)=\int du \  R(u) \exp(\beta u \Sigma) \ ,
\ee
where the normalization condition of $\rho(\s)$ imposes:
\be
\int du \ R(u) \(( 2 \cosh(\beta u)\))^n =1 \ .
\ee
Using this expression for $\rho$,  we obtain in the small $n$ limit:
\be
\ln\((\Tr_{\sigma}\[[\rho(\sigma)^{k+1}\]]\))=
n \int \prod_{i=1}^{k+1}\[[ du_i  R(u_i) \]] 
\ln\(( \[[ \prod_{i=1}^{k+1} {1 \over 2\cosh(\beta u_i)}\]]
\sum_{\s_0}\exp \((\beta \s_0 \sum_i u_i \)) \))
\ee
and:
\bea
&\ln& \(( \Tr_{\sigma,\tau} \[[\rho(\sigma)^{k}\rho(\tau)^{k} 
\exp (\sum_{a=1,n}\beta J \sigma_{a}\tau_{b})\]]\))
= 
n \int \prod_{i=1}^{k}\[[ du_i dv_i R(u_i)R(v_i) \]] 
\\
&& \ \ \ \ \ 
\ln\(( \[[ \prod_{i=1}^{k} {1 \over 4\cosh(\beta u_i)\cosh(\beta v_i)}\]]
\sum_{\s_0,\tau_0} 
\exp
\((\beta \s_0 \sum_{i=1}^k u_i+\beta \tau_0 \sum_{i=1}^k v_i 
+ \beta J_0 \s_0 \tau_0\))
\)) \ .
\eea
Putting these expressions  back into the replica free energy (\ref{repfree}),
one gets exactly the  functional $F[Q]$
which we had written previously in (\ref{bigfree}), provided we
identify the $n \to 0$ limit of
$R(u)$ with the probability distribution (\ref{Rdef}) of the variable $u(J,h)$,
and we use in (\ref{bigfree_det}) a 
function $d(J,h)=c(J,h)[2 \cosh(\beta u(J,h))]$.

In other words we have seen three equivalent ways to solve the Bethe lattice spin glass in the replica symmetric approximation:
\begin{itemize}
\item
One can derive the recursion equations (\ref{RICOR})
for the probability distribution $Q^*(h)$ of the local 'cavity' field $h$,
 and evaluate the free energy and the internal energy 
using this distribution.  
\item
Alternatively one can introduce the free energy functional(\ref{freetot}), which 
depends on the probability distribution $Q(h)$ and satisfies
a variational principle: the distribution $Q^*(h)$ is obtained
as the one which makes the functional stationnary.
\item 
One can obtain the same functional starting from the replica 
approach (\ref{repfree}), making explicitely an assumption of replica symmetry,
 and doing some simple algebra.  It is typical of the replica approach 
that a probability distribution is traded 
with a function of $n$ variables, in the $n \to 0$ limit.
\end{itemize}

\subsection{Free energy shifts}
It may be instructive to compare this approach with the more usual 
cavity method and to check that we obtain the same results.
In  the cavity method, one computes the free energy by 
averaging the various free energy shifts obtained when adding a new site or
a new bond to the lattice.

The first quantity which we compute is the free energy shift
$\Delta F_{iter}$ obtained my adding
a  new spin $\s_0$ connected to $k$ branches, as we do in the iterative procedure.
Using the same notations as in (\ref{iterh}), this free energy shift is:
\be
-\beta \Delta F_{iter}(J_1 \ldots J_{k},h_1 \ldots h_{k})
= \ln\[[2 \cosh \(( \beta \sum_{i=1}^{k} u(J_i,h_i) \)) \]]
 + \sum_{i=1}^{k} \ln\[[ {\cosh (\beta J_i) \over 
\cosh(\beta u(J_i,h_i))} \]]
  \ .
\label{freeshift0}
\ee

In order to compute the total free energy, we also need the
 free energy shift when adding the new spin $\s_0$, onto which
merge $k+1$ branches (see fig. \ref{fig_free}). This free
energy shift is equal to the same quantity with $k$ changed into
$k+1$:
\bea
-\beta \Delta F^{(1)}(J_1 \ldots J_{k+1}, h_1 \ldots h_{k+1})
&=& -\beta F^{(1)} (h_1 \ldots h_{k+1})+
\beta \sum_{i=1}^{k+1} \ln[2 \cosh(\beta h_i)]\\
&=& \ln\[[2 \cosh \((\beta \sum_{i=1}^{k+1} u(J_i,h_i) \)) \]]
 + \sum_{i=1}^{k+1} \ln\[[ {\cosh (\beta J_i) \over 
\cosh(\beta u(J_i,h_i))} \]]
  \ .
\label{freeshift1}
\eea

The free energy shift when adding the two new spins $\s_0,\tau_0$ 
(see fig.\ref{fig_free}) is equal to:
 \be
 \Delta F^{(2)}(J_1 \ldots J_k,K_1\ldots K_k, h_1 \ldots h_k,g_1\ldots g_k )
=F^{(2)}(h_1 \ldots h_k,g_1\ldots g_k )-
 \sum_{i=1}^{k} \ln[4 \cosh(\beta h_i)\cosh(\beta g_i)]
\ee
and is given by:
\bea
-\beta &\Delta& F^{(2)}(J_0,J_1 \ldots J_k,K_1\ldots K_k, h_1 \ldots h_k,g_1\ldots g_k )
=\sum_{i=1}^k \ln \[[ {\cosh( \beta J_i) \over \cosh (\beta u(J_i,h_i))}
 {\cosh( \beta K_i) \over \cosh (\beta u(K_i,g_i))} \]] \\ 
&+&
\ln\[[ \sum_{\s_0,\tau_0} 
\exp\((\beta J_0 \s_0 \tau_0 +\beta\s_0 \sum_{i=1}^k u(J_i,h_i)
 +\beta\tau_0\sum_{i=1}^k u(K_i,g_i) \))
\]] 
\label{freeshift2}
\eea

In the process of adding new sites or new bonds randomly, 
one  can thus compute the total free energy as the average 
over the distribution of fields and couplings of
$[(k+1)/2] \ \Delta F^{(2)} - k \Delta F^{(1)}$. It is a simple exercise 
to check that this indeed gives back the free energy (\ref{bigfree}).
 
\section{The RSB instability}
\label{sect_instab}
The recursion relation of the local fields (\ref{iterh}) 
has been the subject of a lot of 
studies in the past twenty years 
\cite{klein,katsura,nakanishi,bowman,thouless,chayes,mp86,kansom}.  The 
distribution $Q(h)$ is a simple $\delta$ function 
at the origin, indicating a 
paramagnetic phase, at high temperatures $\beta<\beta_c$, 
where the critical inverse 
temperature $\beta_c$ is the solution of\cite{thouless}: 
\be
E_J \tanh^2(\beta_c J) =1/k \ .
\ee   
In the low temperature phase the specific heat becomes negative at low 
enough temperatures at least for some distributions of 
couplings \cite{bowman}, and the 
solution for $Q^(h)$ becomes identical to the replica 
symmetric field distribution 
of the SK model in the large $k$ limit \cite{katsura,bowman,chayes}, 
which is known to be 
wrong.  Another indication that the above procedure gives a wrong 
result for the Bethe 
lattice (while it might be correct for the Cayley tree with some sets of boundary 
conditions \cite{chayes,dewar}) is the fact that 
it fails to identify a transition in a 
magnetic field $H$.  This transition exists though, 
on a line in the $H-T$ plane similar 
to the A-T line, and can be identified by considering 
the onset of correlations between 
two replicas of the system \cite{thouless}.

One can investigate the instability of the previous solution 
using the replica method.  
Writing the recursion relations for the replicated system, 
Mottishaw has shown that the 
replica symmetric solution, which coincides with the simple Bethe-Peierls iteration 
described above, is unstable at $\beta>\beta_c$ 
(or, in a field, beyond the A-T line) 
\cite{motti}.  Unfortunately, getting the replica symmetry 
broken solution in the low temperature phase is difficult.  
In general the problem involves an infinity of order 
parameters which are multi-spin overlaps \cite{orland,matching,motti,motti2,DDGold}.  
As we saw, the replica symmetric solution already involves 
an order parameter which is a whole function 
(the distribution of local fields); going to a `one step RSB' 
solution \cite{MPV}, the 
replica order parameter becomes now a functional, the probability 
distribution over the 
space of local field distributions \cite{monasson98} 
(the reason will be discussed in 
details in the next section).  While one can write formally some integral equations 
satisfied by this order parameter, solving them is in general 
a formidable task.  The 
solution is known only in the neighborhood of the critical 
temperature \cite{motti2}, or in 
the limit of large connectivities \cite{DDGold}, where the 
overlaps involving three spins 
or more are small and can be treated perturbatively, allowing 
for some expansion around 
the SK solution.  In the general case, 
the only tractable method so far has been an 
approximation which parametrizes the functional by a small enough 
number of parameters and 
optimizes the one step RSB free energy inside this subspace \cite{WonShe,BiMoWe}.
We shall develop in the next two sections a solution to this problem.

\section{The formulation of the `one step RSB' solution}
\label{sect_1rsb}
\subsection{The iterative approach}
 In this section we shall explain the physical nature of the RSB
instability and work out the equivalent of the `one step RSB solution' 
using the cavity method 
\cite{cavity,MPV}, i.e. the same type of iterative approach 
which was used in sect.  
\ref{sect_rs_iter}.

The reason for the failure of the simple iterative solution is that it neglects the 
possibility of the existence of several pure states\cite{MPV}.  We shall proceed by 
first assuming some properties of the states on one branch, and then imposing the 
self-consistency of these hypotheses when one joins $k$ branches to a new site.  
Let us 
assume that there exist many pure states, labelled by an index $\alpha$
going from $1$ to $\infty$, with the 
following properties: Looking at one branch of the tree as in fig.1, 
the total local field 
seen by the site $i=0$ at the extremity of this branch depends on the 
state $\alpha$ and 
is denoted by $h_0^\alpha$; the free energies of the states 
on one branch, $F^\alpha$, are 
independent identically distributed (iid) random variables, 
with an exponential density behaving as
\be
\rho(F) = \exp(\beta x (F-F^{R})) \ ,
\label{freedist}
\ee
where $F^{R}$ is a reference free energy. 
The differences between the  free  energies  remain finite when the 
volume goes to infinity, which means that the various states
have non-zero  statistical weights: 
\be
W^{\alpha}= 
{\exp(-\beta F^\alpha) \over \sum_\gamma\exp(-\beta F^\gamma)} \ ,
\label{weight}
\ee
The fact that the $W$ can be normalized in this way is possible 
only if $x<1$.

Let us consider as before a point $i$ on one branch of the Bethe lattice, 
i.e. a point connected to  
$k$ other points.  In each  phase $\alpha$ of the system the magnetization 
$m_{i}^{\alpha}$ will be different and therefore  the effective fields 
$h_{i}^{\alpha}$ depend on $\alpha$.  
The description of the properties of this point will 
include the list of fields $h_{i}^{\alpha}$ and the  free energies 
of the branch $F^{\alpha}$, for all states $\alpha$.  
 Here we shall assume
a relatively simple situation namely that the free energies and 
the magnetic fields are not correlated, and the distribution of free
energies is the one described above, leading to the density (\ref{freedist}).
It is  convenient (to avoid possible
difficulties in dealing with measures in 
infinite dimensional spaces) to order the states in an 
increasing order of free energy, and 
 to consider
only  the set of the first $\cM$ states with lowest free energies (in the end 
$\cM$ will be sent to infinity). Let us introduce the set of the local fields
 in all the $\cM$ states, $\bh=\{ h_i^\alpha \}$. When changing the sample (or
equivalently changing the site $i$), these fields fluctuate and our 
task is to compute the corresponding probability distribution
 $Q(\bh)$, which is a function of the $\cM$ effective 
magnetic fields which is left invariant by the permutations of these fields.
This task is simplified if we assume that the$\cM$ fields $h_i^\alpha$ 
on one point can be characterized as independent random variables.
We thus assume that there exists 
 a probability function $Q_{i}(\bh)$ which can be written in a factorized form:
\be
Q_{i}(\bh)=\prod_{\alpha=1}^\cM Q_{i}(h_{\alpha}) \ .\label{SIMPLE}
\ee
The total probability function $Q(\bh)$ is thus given by 
\be
 Q(\bh)=N^{-1}\sum_{i=1}^N \[[\prod_{\alpha=1}^\cM Q_{i}(h_{\alpha})\]] \ . 
\label{SOMMA}
\ee

In other words on any given  point the fields are independent variables, 
which become correlated on the 
global level after we average over the samples.  
More generally one can  assume that 
the total probability function $Q(\bh)$ is of the form:
\be
Q(\bh)=\int d\lambda \ m(\lambda) \prod_{\alpha=1}^{M} q(h^\alpha|\lambda) \ ,
 \label{QFATT}
\ee
where $\lambda$ is an appropriate set, $m(\lambda)$ 
is a probability distribution, and $ q(h|\lambda)$ is 
a probability distribution on $h$, conditionned to a given value of $\lambda$.
A  possible representation of the distribution (\ref{QFATT})
 is given by  (\ref{SOMMA}), where each point $i$ is 
characterized by a parameter $\lambda$ (extracted with the measure $m(\lambda)$).

We shall now check that this hypothesis is self consistent, i.e. 
that it is reproduced when one iterates the construction of the tree
by merging $k$ lines to a new spin $\s_0$ \cite{general_assump}.
 For each state $\alpha$, the local field on this site 
$h_{0}^\alpha$ is expressed in terms of those on the branches,
 $h_i^\alpha$ by (\ref{iterh}), giving: 
\be
h_{0}^\alpha = \sum_{i=1}^k u(J_i,h_i^\alpha) \ .
\label{iterhRSB}
\ee 
The free energy shift $\Delta F^\alpha$ for the state $\alpha$ 
during this process is 
given by the function $ \Delta F_{iter}$ defined in (\ref{freeshift0}):
\be
\Delta F^\alpha= \Delta F_{iter}(J_1 \ldots J_{k},h_1^\alpha,\ldots,h_k^\alpha)
 \ .
\label{freeshiftRSB}
\ee

  We must be careful at this stage because the free energy shifts and 
the local fields  on the new spin $\s_0$ are correlated.
More precisely, for a given state $\alpha$, $h_{0}^\alpha$ and $\Delta F^\alpha$
are two correlated variables, but they are not correlated 
with the local fields or free-energy shifts in the other states.
Because of our ordering process of the free energies, we need 
to compute and order the new free energies $G^\alpha=F^\alpha+ \Delta F^\alpha$.
$G^\alpha$ and $G^\gamma$ in two different states
  are obviously independent random variables. 
Furthermore, a standard argument of the cavity method \cite{cavity,MPV},
relying on the exponential distribution of the free energies, allows
to show that the new free energy $G^\alpha$ is in
fact {\it  uncorrelated} with the local field $h_{0}^\alpha$.
To show this, let us introduce the joint distribution $P_{0}(h_0,\Delta F)$
of $h_{0}^\alpha$ and $\Delta F^\alpha$. The joint distribution 
$R_{0}(h_0,G)$ of the local field and the new free energy is given
by:
\be
R_{0}(h_{0}, G)\propto \int dF d (\Delta F) \exp(\beta x (F-F^{R}))
P_{0}(h_{0},\Delta F)\delta(G-F-\Delta F)
\propto
\exp (\beta x (G-F^{R})) Q_{0}(h_{0})
\ ,
\ee
where
\be
 Q_{0}(h_{0}) =C \int d (\Delta F) \  P_{0}(h_{0},\Delta F) \exp(-\beta  x\Delta F) 
\label{MAIN}
\ee
the constant $C$ being fixed in such a way that $Q_{0}(h_{0})$ 
is a normalized probability 
distribution.

In our ordering process  of the new free energies $G_\alpha$ we pick up
the $\cM$ lowest ones, sending in the end $\cM$ to infinity. 
 This ordering process thus 
 gives rise to a probability distribution for the $h_{0}^{\alpha}$ 
in which the fields for different $\alpha$ are not  
 correlated and have the distribution $Q_{0}(h_{0})$.
The reader should notice that
this distribution is  in general different from the naive result
$ \int d\Delta F_{0} \  P_{0}(h,\Delta F_{0})$.
In this way we have constructed, for one given new spin $\s_0$ 
with a fixed environment of coupling constants, 
the new distribution of all local fields:
$\prod_\alpha Q_{0}(h_{0}^\alpha)$. By averaging over the coupling
constants, one thus generates 
the probability distribution $Q_{0}(\bh)$ which is  a functional
of the probability distribution $Q(\bh)$ of the other $k$ sites. Imposing that 
\be
Q_{0}(\bh)=Q(\bh) \label{FIXED} \,
\ee
gives  a self-consistency equation for the probability distribution $Q(\bh)$.
We shall see in sect. \ref{sect_algo} how one can  actually find a solution
to this self-consistency equation with a good accuracy. 

Let us suppose for the time being that we know the self-consistent
distribution $Q(\bh)$ and let us evaluate the free energy and  the internal energy.
The computation is quite similar to the one in 
the  replica symmetric case.
There are two contributions to the free energy: the site contribution and the bond 
contribution.  

The local site  contribution to the free energy is evaluated as
a weighted average of the free energy shift when adding one new spin:
\be
F^{(1)}=-{1 \over \beta} E_J \la \ln \((  \sum_{\alpha=1}^\cM  W^\alpha
\exp[-\beta
 \Delta {F^{(1)}}(J_1 \ldots J_{k+1}, h_1^\alpha \ldots h_{k+1}^\alpha)]
 \)) \ra \ ,
\label{fav1}
\ee
where $\Delta{F^{(1)}} $ is the site  free-energy-shift function
 defined in (\ref{freeshift1}), and the bracket denotes an average over 
the  distribution of weights $W^\alpha$
derived from  (\ref{weight},\ref{freedist}),
 and over  the fields (with distribution $ \prod_{i=1}^{k+1} Q(\bh_i)$).

The local bond  contribution to the free energy is evaluated as
a weighted average of the free energy shift when adding a new bond
and the correspoonding two spins:
\be
F^{(2)}=-{1 \over \beta} E_J E_K \la \ln \((  \sum_{\alpha=1}^\cM  W^\alpha
\exp[-\beta 
 \Delta F^{(2)}(J_0,J_1 \ldots J_k,K_1\ldots K_k, 
h_1^\alpha \ldots h_k^\alpha,g_1^\alpha\ldots g_k^\alpha ) ]
 \)) \ra \ ,
\label{fav2}
\ee
where  $\Delta{F^{(2)}} $ is the bond free-energy-shift function
 defined in (\ref{freeshift2}), and the bracket denotes an average over 
the  weights,
 and over  the fields (with distribution 
$ \prod_{i=1}^{k}\[[ Q(\bh_i) Q(\bg_i)\]]$).

The total free energy is given, 
according to (\ref{freetot},\ref{free1},\ref{free2}),
by:
\be
F= {k+1 \over 2} F^{(2)}-k F^{(1)} \ . 
\label{fav}
\ee
Similarly to what happened in the RS case (\ref{bigfree_simple}), 
one can also write
here a simplified form of the free energy, valid only on the saddle point:
\be
F=\frac{k+1}{2}  F^{(1')} -\frac{k-1}{2} F^{(1)} \,
\label{fav_simple} 
\ee
where
\be
F^{(1')}=-{1 \over \beta} E_J \la \ln \((  \sum_{\alpha=1}^\cM  W^\alpha
\exp[-\beta
 \Delta {F^{(1)}}(J_1 \ldots J_{k}, h_1^\alpha \ldots h_{k}^\alpha)]
 \)) \ra \ .
\label{fav1prime}
\ee
For reasons which are beyond our control this second form of the free energy has 
empirically smaller 
errors (about by a factor 3) and less systematic errors (at finite $\cM$) than the 
first one.

The computation of the internal energy is done by considering
what happens when we couple two sites which where previously 
connected each to $k$ branches of the tree. We obtain, with the
same notations as in formula (\ref{bigfree}) and in fig. \ref{fig_free}:
\be
U= -\la   
\sum_{\alpha=1}^\cM \tilde W_2^\alpha 
  J_{0} 
 {\tanh( \beta J_{0}) + \tanh( \beta h_0^\alpha) \tanh( \beta g_{0}^\alpha) 
\over
1+ \tanh (\beta J_{0})\tanh( \beta h_0^\alpha)  \tanh (\beta g_{0}^\alpha)}
\ra 
\label{U1step}
\ee
where $h_0^\alpha= \sum_{i=1}^k u(J_i,h_i^\alpha)$,
$g_0^\alpha= \sum_{i=1}^k u(K_i,g_i^\alpha)$ and we have 
introduced the shorthand notation:
\be
\tilde W_2^\alpha \equiv {
 W^\alpha 
\exp[-\beta \Delta F^{(2)}(J_0,J_1 \ldots J_k,K_1\ldots K_k, 
h_1^\alpha \ldots h_k^\alpha,g_1^\alpha\ldots g_k^\alpha)]
\over
\sum_{\gamma=1}^\cM W^\gamma 
\exp[-\beta \Delta F^{(2)}(J_0,J_1 \ldots J_k,K_1\ldots K_k, 
h_1^\gamma \ldots h_k^\gamma,g_1^\gamma\ldots g_k^\gamma)]
}
\label{W2def}
\ee

The various overlaps can be obtained in the same way. There are now two 
site overlaps, the self-overlap $q_1$ and the inter-state-overlap $q_0$,
which are given by:
\bea
q_1&=& E_J \la {
 \sum_{\alpha=1}^\cM \tilde W_1^\alpha
 \tanh^2[\beta \sum_{i=1}^{k+1} u(J_i,h_i^\alpha)]
}\ra \\
q_0&=& E_J \la {
 \sum_{\alpha \ne \beta} \tilde W_1^\alpha \tilde W_1^\beta
 \tanh[\beta \sum_{i=1}^{k+1} u(J_i,h_i^\alpha)]
\tanh[\beta \sum_{i=1}^{k+1} u(J_i,h_i^\beta)]
}\ra 
\eea
where we have kept  the same notations as in (\ref{fav1}) but we 
have introduced
 $\tilde W_1^\alpha$ which  is a notation for:
\be
\tilde W_1^\alpha \equiv{ 
\exp[-\beta F^\alpha- \beta 
 \Delta {F^{(1)}}(J_1 \ldots J_{k+1}, h_1^\alpha \ldots h_{k+1}^\alpha)]
\over
\sum_\gamma
\exp[-\beta F^\gamma- \beta 
 \Delta {F^{(1)}}(J_1 \ldots J_{k+1}, h_1^\gamma \ldots h_{k+1}^\gamma)]
} \ .
\ee

Similarly, we have two link overlaps $q^{(l)}_1$ and $q^{(l)}_0$ which 
are given by:
\be
q^{(l)}_1= E_J \la {
 \sum_{\alpha=1}^\cM \tilde W_2^\alpha
\(( {\tanh( \beta J_{0}) + \tanh( \beta h_0^\alpha) \tanh( \beta g_{0}^\alpha) 
\over
1+ \tanh (\beta J_{0})\tanh( \beta h_0^\alpha)  \tanh (\beta g_{0}^\alpha)}
\))^2
}\ra 
\label{ql1}
\ee
and
\bea
q^{(l)}_0=& E_J& 
\la 
 \sum_{\alpha \ne \beta} \tilde W_2^\alpha \tilde W_2^\beta
\(( {\tanh( \beta J_{0}) + \tanh( \beta h_0^\alpha) \tanh( \beta g_{0}^\alpha) 
\over
1+ \tanh (\beta J_{0})\tanh( \beta h_0^\alpha)  \tanh (\beta g_{0}^\alpha)}
\)) \\
&&\ \ \ \ \ 
\(( {\tanh( \beta J_{0}) + \tanh( \beta h_0^\beta) \tanh( \beta g_{0}^\beta) 
\over
1+ \tanh (\beta J_{0})\tanh( \beta h_0^\beta)  \tanh (\beta g_{0}^\beta)}
\)) 
\ra 
\label{ql0}
\eea
where we keep the same notations as in (\ref{U1step},\ref{W2def}).

At this stage we have written the hole formalism of the cavity 
method at the level of one step RSB. The self-consistency equation
(\ref{FIXED}) fixes the distribution $Q(\bh)$, from which one can deduce
the free energy and internal energy through (\ref{fav}) and (\ref{U1step}).
One can actually find a self-consistent solution for any value of
the parameter $x \in [0,1]$, and the free energy depends on $x$
through the distribution of free energies. 
It is well known that, in order to describe the thermal equilibrium,
one must fix $x$ 
 by maximising the free energy with respect to $x$ 
\cite{MPV,monasson,miguel}.  (Actually 
the whole dependance on $x$ carries some information \cite{monasson,pot,miguel}, 
particularly interesting for optimization problems, which we shall not try to 
study here since this paper is restricted to the 
study of equilibrium thermodynamics).
As we shall see, the 
variation of free energy with respect to $x$ is small and one needs a 
better computation 
of the derivative than just a naive difference.

We have improved the precision on the computation of the free energy and its 
$x$ derivative by using the  theorem of appendix 1, which allows to 
compute
explicitely the derivative of the free energy with respect to $x$. 
From the structure of \form{fav} one finds that the total derivative
$d=dF/dx$ takes the form :
\be
d(x)= -{1 \over x} F - k d^{(1)} +{k+1 \over 2} d^{(2)} \ ,
\label{derivative}
\ee
where 
\be
d^{(1)}=-{1 \over x} E_J \la \ln \((  \sum_{\alpha=1}^\cM  W^\alpha
\exp[-\beta
 \Delta {F^{(1)}}(J_1 \ldots J_{k+1}, h_1^\alpha \ldots h_{k+1}^\alpha)]
\ln[ \Delta {F^{(1)}}(J_1 \ldots J_{k+1}, h_1^\alpha \ldots h_{k+1}^\alpha)]
 \)) \ra \ ,
\label{d1}
\ee
using the notations of (\ref{fav1}), and 
\be
d^{(2)}=,
-{1 \over x} E_J E_K \la \ln \((  \sum_{\alpha=1}^\cM  W^\alpha
\exp[-\beta 
 \Delta F^{(2)}(J_0,J_1 \ldots J_k,K_1\ldots K_k, 
h_1^\alpha \ldots h_k^\alpha,g_1^\alpha\ldots g_k^\alpha ) ] 
\ln[\Delta F^{(2)}]
 \)) \ra \ ,
\label{d2}
\ee
using the notations of (\ref{fav2}).

\subsection{A variational formulation}
As in the case where no replica symmetry breaking is present 
we can write a free energy functional of the field 
distribution $Q(\bh)$ such that the self-consistency equations 
for $Q$ are equivalent to the stationnarity condition of this functional. 

This free energy functional is a simple 
generalisation of the replica symmetric one, given by:
\bea
{F[Q] \over N}=&&\frac{k+1}{2}  \int \prod_{i=1}^k
\[[ d\bh_i d\bg_i Q(\bh_{i})Q(\bg_{i}) \]]
 \ 
F^{(2)}(\bh_1 \ldots \bh_k,\bg_1\ldots \bg_k ) \\
&-& k \int \prod_{i=1}^{k+1}\[[ d\bh_i  Q(\bh_{i}) \]] \ 
 F^{(1)}(\bh_1\ldots \bh_{k+1} ) \, 
\label{bigfree_rsb}
\eea
where
\be
-\beta F^{(1)}(\bh_1\ldots \bh_{k+1})=E_J \la
\ln\(( \sum_\alpha W^\alpha \[[\prod_{i=1}^{k+1} 
{1 \over 2\cosh(\beta h_i^\alpha)} \]]
\sum_{\s_0,\s_1,\ldots \s_{k+1}}
\exp\[[\beta \s_0 \sum_{i=1}^{k+1}J_i \s_i+\beta \sum_{i=1}^{k+1}h_i^\alpha \s_i\]]
 \)) \ra \ ,
\label{bigfree1_rsb}
\ee
\bea 
-\beta F^{(2)}(\bh_1 \ldots \bh_k,&\bg_1&\ldots \bg_k)=E_J E_K \la
\ln\(( \sum_\alpha W^\alpha \[[\prod_{i=1}^{k} 
{1 \over 4\cosh(\beta h_i^\alpha) \cosh(\beta g_i^\alpha)} \]]
\sum_{\s_0,\s_1,\ldots \s_{k}} \sum_{\tau_0,\tau_1,\ldots \tau_{k}} \right.
\\
&&
\left.
\exp\[[\beta J_0 \s_0 \tau_0 +\beta\s_0 \sum_i J_i \s_i
+ \beta \sum_i h_i^\alpha \s_i 
+\beta\tau_0\sum_i K_i \tau_i+\beta\sum_i g_i^\alpha \tau_i \]]
\)) \ra \ .
\label{bigfree2_rsb}
\eea
The weights $W_\alpha$ are given by (\ref{weight}), and the brackets stand for
an average over the distribution of free energies (\ref{freedist}).

 The proof of the equivalence of the stationarity equation of this functional
 with the self-consistency condition of the iterative procedure
can be done exactly as in the replica symmetric case.
The advantage of  this variational formulation is that 
we can compute directly  the various derivatives of 
the free energy (e.g. with respect to $x$  and with respect to $\beta$) 
by taking into account only  the explicit dependence.

\subsection{Equivalence with the  replica formalism}
We can compare what we have obtained in the previous sections with 
the results from the replica formalism.
In the one step RSB  formalism the $n$ replicas are divided into  $n/x$ 
groups (labeled by $\cC$) of $x$ replicas each, and  
 the function $\rho(\sigma)$ depends on the $n/x$ `block' variables
\be
\Sigma_{\cC}=\sum_{a\in \cC} \sigma_{a}\ ,
\ee
each sum containing $x$ terms.

In this case we face the problem that $\rho(\sigma)$ may depend on 
the $n/x$ variables $\Sigma_{\cC}$
in a rather complex way. Similarly  to what we have done
in the iterative approach, we shall not try to describe the most 
general dependence,
 but we restrict to the class of probability distributions  $\rho(\sigma)$
which can be written as: 
\be
\rho(\sigma)=\int d\lambda \mu(\lambda) \int\prod_{\cC=1}^{n/x} \[[ du_{\cC}
\phi (u_{\cC}|\lambda) 
\]]
\exp\((\beta \sum_{\cC=1}^{n/x}u_{\cC} \Sigma_{\cC}\)) \ . 
\label{VOILA}
\ee
with a positive probability distribution $\mu(\lambda)$
\cite{WonShe} and a positive function $\phi(u|\lambda)$. 
The normalisation condition
on  $\rho(\sigma)$ is implemented by imposing that:
\be
\forall \lambda: \ \ \ 
\int du \phi(u|\lambda) (2 \cosh(\beta u))^x =1 \ ,
\label{norm_P}
\ee
so that the function
\be
\Phi(u|\lambda)= \phi(u|\lambda) (2 \cosh(\beta u))^x
\ee
is a probability distribution on the $u$ variable, 
for any value of $\lambda$.

It is easy to check that this form for the function is consistent with 
 the stationarity equations for the free energy (\ref{rhoeq}) by proving
that, if $\rho(\sigma)$ has the form (\ref{VOILA}), so does
\be
 E_J{
\Tr_{\tau} \((\rho(\tau)^{k} 
\exp [\sum_{a=1}^{n}\beta J \sigma_{a}\tau_{a}]  \)) \ .
}
\ee

Using the Ansatz (\ref{VOILA}) for $\rho$, one can write the
`site' term in the replica free energy (\ref{repfree}) as:
\be
\Tr_{\sigma}\((\rho(\sigma)^{k+1}\))=\int \[[\prod_{i=1}^{k+1}d\lambda_{i}
\mu(\lambda_{i})\]] A\((\lambda_1,\ldots,\lambda_{k+1}\))^{n/x} \ ,
\ee
where:
\be
 A\((\lambda_1,\ldots,\lambda_{k+1}\))=
 \int\prod_{i=1}^{k+1} \[[ du_i \Phi(u_{i}|\lambda_{i}) \]]
\[[ {2\cosh(\beta \sum_{i=1}^{k+1 } u_{i}) 
\over 
\prod_i \[[2 \cosh \beta u_i\]] }
\]]^{x}
  \ .
\ee

Using the small theorem proven in  appendix I, 
the previous expression can be written as
\be
 \ln A\((\lambda_1,\ldots,\lambda_{k+1}\))= x
 \int\prod_{i=1}^{k+1} \prod_\alpha \[[du_i^\alpha
 \Phi(u_{i}^\alpha|\lambda_{i}) \]]
  \la  \[[ \ln  \((\sum_\alpha W_\alpha 
{2\cosh(\beta \sum_{i=1}^{k+1 } u_{i}^\alpha) 
\over 
\prod_i  \[[2 \cosh \beta u_i^\alpha \]] } \)) \]] \ra \ ,
\ee
where the bracket means an average over the
distribution of the weights $W_\alpha$ have the usual distribution of the weights over
states, parametrized by the parameter $x$ (see appendix I).
One gets in the end:
\bea
 \ln&& \Tr_{\sigma}\((\rho(\sigma)^{k+1}\))= n \int \prod_{i=1}^{k+1}
\[[d\lambda_{i}
\mu(\lambda_{i})\]] \\
&&\int\prod_{i=1}^{k+1} \prod_\alpha \[[du_i^\alpha
 \Phi(u_{i}^\alpha|\lambda_{i}) \]]
\la
 \ln\(( \sum_\alpha W_\alpha 
{2\cosh(\beta \sum_{i=1}^{k+1 } u_{i}^\alpha) 
\over 
\prod_i\[[ 2 \cosh \beta u_i^\alpha \]] } \))
\ra
\eea

The previous quantity is exactly the expression of the `site contribution'
to the variational 
free energy found in (\ref{bigfree1_rsb}), provided we identify
the  probability distributions defined in the 
iterative approach (\ref{QFATT})
and in the replica approach (\ref{VOILA}):
\be
\mu(\lambda)=m(\lambda) \ \ ; \ \
 \Phi(u|\lambda)= E_J \int dh \  q(h|\lambda) \delta(u-u(J,h)) \ .
\ee
 A similar computation shows that the `bond' term in the free
energy, calculated either through the iterative procedure or through
the replica method also coincide.

We have thus derived the variational equations 
for the probability distribution of local fields
 in the one step RSB case using  two different methods, 
the cavity iterative approach on the one hand, and the
 algebraic replica formalism on the other hand.

\section{Solving the one step RSB equations}
\label{sect_algo}

Our method consists in following the population of local fields $h_i^\alpha$ when
one iterates the merging process of $k$ branches onto one
site. In some sense it thus amounts to
solving the complicated equation for the functional order parameter
by a method of population dynamics. In other words 
we parametrize the probability 
distribution by presenting a large number of instances of variables 
distributed according to this probability distribution.
A similar method 
has ben first used in the context of mean field equations  in \cite{ANTH}.

To explain it in more details,
let us first state how the procedure works in the case of the `replica symmetric'
approximation of sect. \ref{sect_rs}. There, one just chooses a population of 
$\cN$ local fields $h_i$. At each iteration, one picks up $k$ such fields at random
among the $\cN$, and computes the new field $h_0$ according to (\ref{iterh}). Then
one field is removed at random from the population and is substituted by $h_0$.

In this way one defines a Markov chain on the space 
of the $\cN$ magnetic fields. This 
chain has a stationary distribution which is reached after some transient time.
In the 
 $\cN \to \infty$ limit, the stationary distribution 
satisfies the self-consistency equation(\ref{RICOR}). It is possible 
to argue that the corrections to this limit are proportional to
$\cN^{-1}$ and could also be computed analytically. Our procedure 
consists in fixing the value of $\cN$, iterating the merging transformation 
many times in such a way as to  obtain the average 
over the asymptotic distribution at fixed $\cN$  with a high precision,
 and  finally  extrapolating the results to  $\cN \to \infty$.

If we consider the case where there exist  many states,
we have the same problem as before, with 
the only difference that at each point we have a probability 
distribution $Q_{i}(\bh)$. We must therefore consider a 
population dynamics in which the elements of the population 
are probability distributions.
We use the population method to represent the probability distribution 
in each point $i$ by a populations of fields. In this way we have 
a population of $\cN$ 
populations of $\cM$ fields (a total of $\cN \cM$ fields
 $h_i^\alpha$, $i \in \{ 1,...,\cN \}$,
$\alpha \in \{ 1,...,\cM \}$), where both $\cN$ and $\cM$ have to 
go to infinity.

The basic step of the algorithm is the merging of $k$ lines. One
chooses $k$ sites $i_1,...,i_k$ in $\{ 1,...,\cN \} $, and one
generates, for each of the  $ \cM$ states, a new local field
$h_0^{\alpha}$ obtained by merging $k$ branches, using
(\ref{iterh}), as well as the corresponding free-energy shift
 $\Delta F^{\alpha}$ calculated using (\ref{freeshift0}). But the
 field $h_0^{\alpha}$ is not the one which will enter the 
population of fields. The reason is that one needs to reweigh
the various states by a factor which depends on their free energy
shifts: as seen in (\ref{MAIN}), the field distribution,
at a fixed new free energy, is modified by a factor $exp(-\beta x \Delta F)$.
 From the knowledge of the $h_0^{\alpha}$ one can infer an approximate 
form of the distribution $P_0(h_0)$ from which they are extracted.
For instance a simple form for $P_0$ is a smoothly interpolated  version of the 
identity 
\be
\int_{-\infty}^h P_0(h_0) dh_0 = (1/\cM) \sum_\alpha \theta (h_0-h_0^{\alpha})
\ee
where $\theta(x)$ is Heaviside's function (in practice we  smooth this
staircase function by a linear interpolation procedure). According to
(\ref{MAIN}), the real field distribution $Q_0(h)$ is well approximated by
 a smoothly interpolated  version of the 
identity 
\be
\int_{-\infty}^h Q_0(h) dh = (1/\cM) \sum_\alpha 
\exp(-\beta x \Delta F^{\alpha}) \theta (h-h_0^{\alpha}) \ .
\label{reweiB}
\ee

We can use two different methods in order to achieve the reweighing, 
which will lead to two different algorithms.
\begin{itemize}
\item Method A: 
The idea is to generate, from the set of $\cM$ fields $h_{i_l}^{\alpha}$, 
on each of the sites $i_1,...,i_k$ which is used in the merging,
a {\it larger} population, of $r \cM$ local fields
$h_{i_l}^{\alpha'}$, $\alpha' =1,...,r \cM$, taken from the same distribution.
This will be realised by having
 \be
(1/\cM) \sum_{\alpha=1}^\cM  \theta (h-h_{i_l}^{\alpha})
\simeq
(1/(r\cM)) \sum_{\alpha'=1}^{r \cM} 
 \theta (h-h_{i_l}^{\alpha'}) \ ,
\label{reweightA}
\ee
at the level of linearly interpolated functions.
Simultaneously one 
generates $r \cM$ independent random free energies $F^{\alpha'}$,
$\alpha' =1,...,r \cM$, with the exponential density (\ref{freedist}).
 For each of the  $r \cM$ states, one then 
compute the  local field $h_0^{\alpha'}$  and the 
free energy shift $\Delta F^{\alpha'}$. The correct reweighing is obtained
by the selection of low lying states: one computes the new  $r \cM$  free energies 
$F^{\alpha'}+\Delta F^{\alpha'}$, orders them, and keeps only the $\cM$ states with
the lowest new free energies. Their local fields,
called  $h^{\alpha}$, with $\alpha \in \{ 1, \cdots, \cM \}$,
have the correctly reweighed distribution
 provided that $r$ is large enough
so that there is a zero probability for the states $\alpha'$ with the highest
free energy $F^{\alpha'}$ to enter the list of the $\cM$ selected states
after reweighing.

\item Method B:
The idea is to generate directly the fields with the reweighed
distribution (\ref{reweiB}).
Knowing the
$\cM$  local fields  $h_0^{\alpha}$ and their free-energy shifts
$\Delta F^{\alpha}$, we generate $\cM$ new  local fields  $h^{\alpha}$
 in such a way that
the following identity holds at the level of linear interpolation:
\be
{1 \over\sum_\alpha \exp(-\beta x \Delta F^{\alpha})} \sum_\alpha 
\exp(-\beta x \Delta F^{\alpha}) \theta (h-h_0^{\alpha})
\simeq
{1 \over \cM} \sum_\alpha 
 \theta (h-h^{\alpha}) \ .
\label{reweightB}
\ee
\end{itemize}
Having generated the new fields $h^{\alpha}$ which are typical of the
properly reweighed distribution,
we then substitute in the population
the set of $\cM$ local fields $h_i^\alpha$, $\alpha=1,..,\cM$
by the  set of new fields $h^{\alpha}$. The site index $i$ on which this
substitution is performed is chosen sequentially.

While the merging of $k$ lines is enough to build up the Markov 
chain which generates the population of local fields, one also needs 
to consider some different merging
processes in order to compute the various observables, free energy,
energy, local overlap and link overlap.

By merging $k+1$ lines instead of $k$, one generates with exactly the 
same procedure as above the three sets of $\cM$ local fields
$h_0^\alpha$, free-energy shifts
$\Delta F^{\alpha}$ and new fields $h^{\alpha}$ (we call new fields the
fields which are typical of the reweighed distribution,
obtained either through procedure A or B). Using the little
theorem of Appendix I, the site contribution (\ref{fav1}) 
to the free energy is computed as
\be
F^{(1)}= -(1/(\beta x)) \ln \[[ (1/\cM) 
\sum_\alpha \exp(-\beta x \Delta F^{\alpha} )\]] \ ,
\label{free1algo}
\ee
 and the  site
overlaps receive a contribution 
\be
q_1= \sum_\alpha \tanh^2 (\beta h^{\alpha}) \ \ \ ; \ \ \ 
q_0= \sum_{\alpha \ne \gamma} \tanh (\beta h^{\alpha})
\tanh (\beta h^{\gamma}) \ .
\label{ovsitealgo}
\ee
The $x-$ derivative of the free energy (\ref{d1}) 
receives a site contribution equal to:
\be
d^{(1)}= -(1/(\beta x)) \ln \[[ (1/\cM) 
\sum_\alpha \exp(-\beta x \Delta F^{\alpha} ) \ln [ \Delta F^{\alpha}] \]] \ .
\label{d1algo}
\ee
These contributions are then averaged over many iterations.

One can also add 	a new link to the system. This is done by merging
$k$ lines on the left side of the new link, generating the local fields 
$h_0^\alpha$, and similarly merging $k$ lines on the right side of the 
new link, generating the local fields $g_0^\alpha$
$h_0^\alpha$. The corresponding free energy shift  $\Delta F^{\alpha}$
is now computed using (\ref{freeshift2}). The bond contribution 
(\ref{fav2}) 
to the free energy is computed by the same expression
as (\ref{free1algo}), but using this free energy shift 
$\Delta F^{\alpha}$ corresponding
to a bond-addition.

The $x-$ derivative of the free energy (\ref{d2}) 
receives a bond contribution (\ref{d2}) equal to:
\be
d^{(2)}= -(1/(\beta x)) \ln \[[ (1/\cM) 
\sum_\alpha \exp(-\beta x \Delta F^{\alpha} )\ln [ \Delta F^{\alpha}]\]] \ .
\label{d2algo}
\ee
 In order to compute the
internal energy and the link overlaps, it is useful to introduce 
the local fields $v_0^\alpha$ defined by
\be
v_0^\alpha= {1 \over \beta} \atanh
 {\tanh( \beta J_{0}) + \tanh( \beta h_0^\alpha) \tanh( \beta g_{0}^\alpha) 
\over
1+ \tanh (\beta J_{0})\tanh( \beta h_0^\alpha)  \tanh (\beta g_{0}^\alpha)} \ .
\label{vdef}
\ee
From this population one generates a population of new fields $v^\alpha$
which takes into account the appropriate weights of the fields
with one of the two procedures A or B, similarly
to what was done in (\ref{reweightA}) or in (\ref{reweightB}) 
for going from the fields $h_0^\alpha$
to $h^\alpha$.
From (\ref{U1step},\ref{ql1},\ref{ql0}), the contributions to the
internal energy and link overlaps are given by:
\be U=- J_0 \sum_\alpha
\tanh( \beta v^\alpha) \ ,
\label{Ualgo}
\ee
\be
  q_1^{(l)}= \sum_\alpha \tanh^2 (\beta v^{\alpha}) \ \ \ , \ \ \ 
q_0^{(l)}= \sum_{\alpha \ne \gamma} \tanh (\beta v^{\alpha})
\tanh (\beta v^{\gamma}) \ .
\label{ovlinkalgo}
\ee
In order to compute the free energy with the simplified formula 
(\ref{fav_simple}), one also needs the contribution $F^{(1')}$
to the free energy, which is obtained by the same expression
as
 (\ref{free1algo}), but using the free energy shift 
$\Delta F^{\alpha}$ obtained by merging $k$ lines.

Let us therefore summarize the main lines of our
population dynamics algorithms for solving the Bethe lattice spin-glass
problem at the level of one-step RSB. We have used two algorithms,
A and B, which differ in the reweighing procedure used,
but have otherwise the same skeleton:

\begin{enumerate}
\item  Start from the population of $\cN \times \cM$ local fields  $h_i^\alpha$.
\item Merge $k+1$ lines and compute the site observables:
  \subitem a)  Choose at random $k+1$ sites $i_1,...,i_{k+1}$ 
  in $\{ 1, ...,\cN \} $.
  \subitem b) For each of these $k+1$ sites, say on site $j \in \{ i_1,...,i_{k+1}
   \}$,
  one has a population of $\cM$ local fields $h_j^\alpha$, $\alpha =1,...,\cM$.
  For each of the  $ \cM$ states,
  compute the  new local field $H_0^{\alpha}$ obtained by
  merging $k+1$ branches, using
  (\ref{iterhkp1}). Compute  the corresponding
  free energy shift $\Delta F^{\alpha}$ using (\ref{freeshift1}).
  \subitem c)  Knowing the sets of fields $H_0^{\alpha}$ and
  free energy shift $\Delta F^{\alpha}$, generate a new set
  of fields $H^{\alpha}$ according to (\ref{reweightA}) 
  in algorithm A (resp.(\ref{reweightB}) 
  if one uses algorithm B). 
  \subitem d) Compute the site contribution to the free energy using
  (\ref{free1algo}), its $x-$derivative using (\ref{d1algo})
   and the contribution to the site overlaps 
  (\ref{ovsitealgo}).
\item Merge $2 k$ lines onto a new bond and compute the bond overlaps:
  \subitem a)  Choose at random $2 k$ sites 
  $i_1,...,i_{k},j_1,...,j_{k}$ in $\{ 1, ...,\cN \} $.
  \subitem b) From the  sites $i_1,...,i_{k}$,
  compute the $\cM$  local fields $h_0^{\alpha}$ obtained by
  merging  the $k$ branches, using
  (\ref{iterh}). From the  sites $j_1,...,j_{k}$,
  compute the $\cM$  local fields $g_0^{\alpha}$ obtained by
  merging  the $k$ branches, using
  (\ref{iterh}). Deduce the $\cM$  local fields $v_0^\alpha$ 
  using (\ref{vdef}). 
	\subitem c) Compute  the 
  free energy shifts $\Delta F^{\alpha}$ using (\ref{freeshift2}).
	\subitem d)  Knowing the sets of fields $v_0^{\alpha}$ and
  free energy shifts $\Delta F^{\alpha}$, generate a new set
  of fields $v^{\alpha}$ according to (\ref{reweightA}) 
  in algorithm A (resp.(\ref{reweightB}) 
  if one uses algorithm B).
	\subitem e) Compute the bond contribution to the free energy using
  (\ref{free1algo}), its $x-$derivative using (\ref{d2algo}) and 
  the contribution to the
	internal energy (\ref{Ualgo}) and link overlaps 
  (\ref{ovlinkalgo}).
\item Merge $k$ lines and update the population of fields:
  \subitem a)  Choose at random $k$ sites $i_1,...,i_k$ in $\{ 1,...,\cN \} $.
  \subitem b)  For each of these $k$ sites, say on site $j \in \{ i_1,...,i_k   \}$,
  one has a population of $\cM$ local fields $h_j^\alpha$, $\alpha =1,...,\cM$.
  For each of the  $ \cM$ states,
  compute the  new local field $h_0^{\alpha}$ obtained by
  merging $k$ branches, using
  (\ref{iterh}). Compute  the corresponding
  free energy shift $\Delta F^{\alpha}$ using (\ref{freeshift0}).
  \subitem c) Knowing the sets of fields $h_0^{\alpha}$ and
  free energy shift $\Delta F^{\alpha}$, generate a new set
  of fields $h^{\alpha}$ according to (\ref{reweightA}) 
  in algorithm A (resp.(\ref{reweightB}) 
  if one uses algorithm B).
  \subitem d) Compute  the contribution $F^{(1')}$ to the simple
  form (\ref{fav_simple}) of the free energy using
  (\ref{free1algo}).
  \subitem e)  Pick up  the  site $i\in \{ 1,..., \cN \}$ sequentially, 
  and substitute the 
  local fields $h_i^1,....,h_i^\cM$ by the new  local fields $h^{\alpha}$.
\item  Start again the iteration from 2).
\end{enumerate}
 
Obviously one needs not really perform the above three merging
procedures sequentially. In our actual algorithm, we select $2k+2$
random points, merge two groups of $k+1$ to compute site ovbservable,
two groups of $k$ to compute bond observables and to update two new sets
of $\cM$ fields.
 
A word about the difference between the two algorithms. In 
algorithm A,
the value of $r$ must be chosen large enough so that the probability 
of a state with the highest old free energy $F^{\alpha'}$ to 
enter the set of selected 
$\cM$ states be negligible.  Both $\cM$ and $r$ must go to 
infinity and in the numerical computations we have taken $r=\cM$.
 Algorithm B is   faster than  algorithm (A) by a 
factor that is asymptotically proportional to $r$ for large $r$.  
In algorithm B
there is no need of introducing $r \cM$ fields at an intermediate 
stage: it corresponds to 
the discretization of eq.  \ref{MAIN} and we care take of the 
effects of the reshuffling 
by explicitly reweighing the fields.  Unfortunately,
as we shall see in the next section, the finite 
$\cM$ corrections are 
empirically  larger in algorithm B that in
algorithm A. In our case algorithm B
turned out to be faster by a  factor about 10, but
we mentioned both algorithms because algorithm A is somewhat 
simpler conceptually (and closer to the original discussion
of the cavity method), and also because in different situations 
(e.g.depending on the value 
of $x$) the relative advantages may be reversed.

\section{Numerical results}
\label{numerics}
Here we present the numerical results for one case in order 
to study the dependence of 
the algorithm on the various parameters involved in the numerical computation.

We consider the Ising spin glasses with binary couplings ($J=\pm 1$) 
on a random lattice 
of fixed coordination 6 (k=5). High precisions measurements \cite{MAZU} 
of the internal energy are
available at temperatures greater or equal to .8 for different 
values of the number of spins $N$ (up to 
$N=4096$). The data of the energy at $T=.8$ versus
size can be very well fitted by a power law correction:
\be
U(N)=U+A N^{-\omega}
\ee
with a quite reasonable value of $\omega \sim .767 \pm .008$ and $A \sim 2.59 \pm
.02$.
 The value of the internal energy
at infinite size is  estimated to be
\be
U= -1.799 \pm .001   \ .
\label{res_simu}
\ee

We have done a replica symmetric computation for different values of
the population size $\cN$. 
For large $\cN$ there are corrections proportional to $1/N$ 
(as expected) and  for $\cN > 10^{3}$ the $1/\cN$ corrections are 
negligible within our accuracy. With $I=100,S=1000,\cN=4000$,
we obtain the following replica symmetric results for the free energy, internal
energy, entropy, site and link Edwards-Anderson parameters:
\bea
F=-1.863\pm 0.002 \ \ , \ \ U=-1.8160 \pm  0.001 \ \ , \ \ 
S=.058 \pm 0.004
\\
q=0.6863\pm  0.0002  \ \ , \ \ q_{link}=	0.6385 \pm  0.0003 \  .
\label{res_rs}
\eea
Notice that the value of the internal energy totally 
 disagrees with the one found in the simulations
(\ref{res_simu}).

\begin{figure}
\hbox{\epsfig{figure=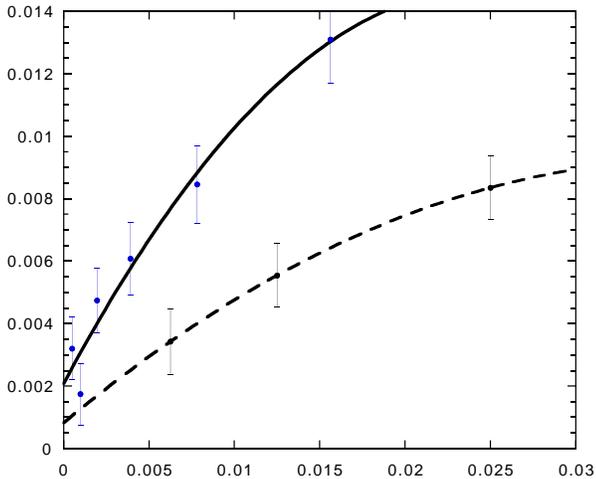,width=8cm
}}
\caption{The $x$-derivative of the free energy, $d(x)$,
 evaluated at $x=.21$, is plotted  as function of $\cM^{-1}$ for the 
algorithm $A$ (upper curve) and the algorithm $B$ (lower curve). }
\label{EXTRA}
\end{figure}
We have done a computation at the one step RSB level using the two
algorithms described in the previous section, always at temperature $T=.8$.

The crucial point is to find the value of the parameter $x$ such that 
the derivative of the  free energy with respect to $x$ vanishes.
In fig.\ref{EXTRA} we show 
our results  for the derivative $d(x)$ at $x=.21$, plotted
versus  different values of $\cM$. 
We have used  both  algorithms A and  
B. With algorithm B we have used $\cM=2^{l},l=3\ldots 12$
and we plot the results obtained for $l\le 10$ for a clearer figure.
With algorithm A we have used $r=\cM \le 400$.
Unfortunately
the finite $\cM$ corrections are 
empirically much larger in the a-priori-faster algorithm B.
  Moreover, in this case
although the finite $\cM$ corrections seem to be 
asymptotically proportional to $1/\cM$, 
high order corrections cannot be neglected unless $\cM$ is very large.
In the end both algorithms give compatible asymptotic results at large $\cM$
as seen on fig.\ref{EXTRA},
with similar computer efforts (for this temperature and the values of
$x$ which are relevant).

\begin{figure}
\hbox{\epsfig{figure=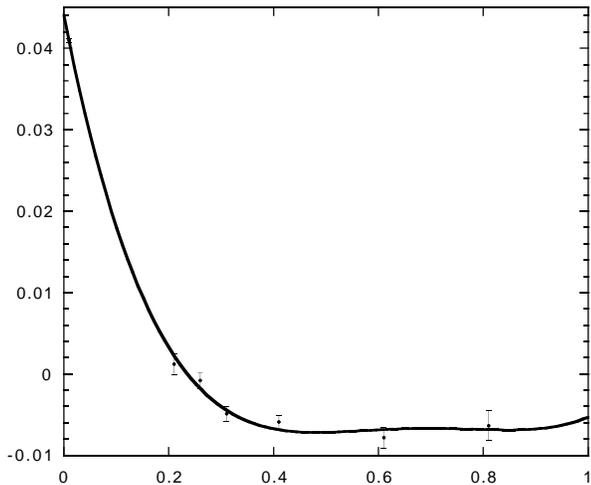,width=8cm
}}
\caption{The $x$-derivative of the free energy , $d(x)$,
 computed using the algorithm (B) 
and extrapolated at large values of $\cM$,
 plotted versus $x$.
}
\label{FINAL}
\end{figure}

In fig.\ref{FINAL} we  plot the extrapolated result at  $\cM=\infty$ for
the free energy derivative $d(x)$, obtained using 
algorithm $B$.
The  data has been extrapolated with a second order polynomial of $\cM^{-1}$ in 
the interval $\cM=[256-4096]$.

Replica symmetry breaking is clearly present. The value of $x$
where the free energy is maximum, which can be obtained by 
estimating the zero of the function $d(x)$, is given by  
$x^{*}= .24\pm .02$. We use a similar procedure for all other observables:
we extrapolate the $x$ dependent results at $\cM=\infty$ and evaluate the 
errors due to the imprecise location of $x^{*}$ (which is by far the 
largest source of 
error). This gives the following values of the free energy, 
internal energy, entropy,
site and link overlaps:
\bea
F=-1.858\pm .002 \ \ , \ \ U=-1.799 \pm  0.001 \ \ , \ \ 
S=.074 \pm 0.004
\\ 
q_1=0.779\pm.006  \ \ , q_0= .30 \pm .02
\\
\ \ q_1^{(l)}=.706	\pm.007  \ \ , q_0^{(l)}=.408 \pm .01
\label{res_rsb}
\eea

The energy is in very good agreement with the results from simulations. 
In order to compare the values of the overlaps, one can study
the quantity $\langle q^{2}\rangle= \int dq P(q) q^{2}$. In our
RSB theory  we find $\langle 
q^{2}\rangle=(1-x^*)q_1^2+x^* q_0^2=.485\pm.015$ which agrees
well with   the numerical value $\langle 
q^{2}\rangle=.49\pm.02$. 
In order to perform a finer  comparison 
 it is useful to consider a quantity which 
is sensitive to replica symmetry breaking. A  natural choice is
\be
R= \int dq P(q) q^{4} -\left(\int dq P(q) q^{2}\right)^{2}
\ee
We find in our RSB theory$R=.046\pm .002$ which is 
again in good agreement with the result of the simulations 
extrapolated at infinite volume: $R=.051\pm.002$. 
The agreement is remarkable if 
we recall that in the replica symmetric case $R=0$. 
The possible small difference between our 
 value  and the simulation data is likely due to the
the approximation of one step replica symmetry breaking 
(it is quite likely that replica 
symmetry should be broken in a continous way, as happens in the limit of infinite 
coordination number).
 One should notice that the order parameter $q_0$ is non zero, 
which explains why some previous attempts at solving the one step RSB problem within
a restricted subspace with $q_0=0$ did not improve much on the RS solution
\cite{WonShe} (the necessity of having a non vanishing
$q_0$ was already noticed in \cite{goldschmidt_gpref}).

\begin{figure}
\hbox{\epsfig{figure=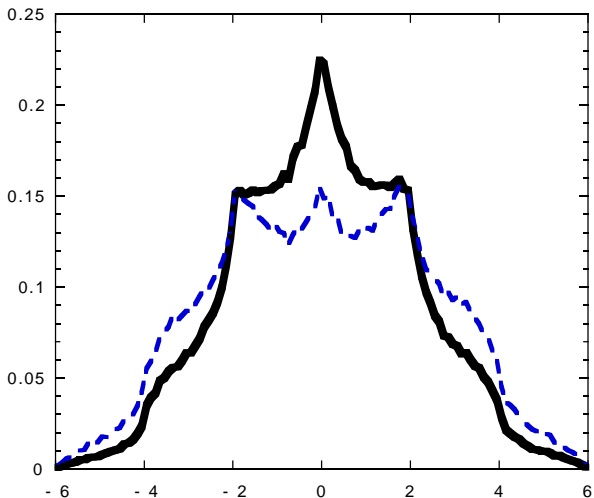,width=8cm}} 
\caption{The probability distribution of the field $H_{0}$ at 
$x=1$ before (full curve) 
and after the reweighing (dashed curve).}
\label{CUM}
\end{figure}

Finally let us point out the crucial effect of the reweighing of the states which
modifies the flocal fields as in (\ref{MAIN}). In fig.  (\ref{CUM})
we plot the probability distribution of the field 
$H_{0}$ before and after the reweighing, at $x=1$.

\section{Conclusion}
\label{conclusion}
We have presented a general solution of finite connectivity 
spin glass problems at the level of one step RSB. This solution
uses the cavity method, together with a kind of population dynamic algorithm
to solve the complicated functional equations. As exemplified by a 
detailed numerical study of the sping glass on a random lattice with fixed
connectivity, this method aloows to obtain good agreement with
numerical simulations.
It should be rather easily extendable to many other disordered 
systems with a finite 
connectivity, including the fluctuating valence spin glass and various optimization 
problems such as the K-satisfiability problem. 

The possibility to go to higher order of  RSB should be explored. In
principle the method we have presented can be extended to higher order.
At second order one will need a population of $\cM_s$ states within a given
cluster, and a population of $\cM_c$ clusters of states. ITherefore
the algorithm must follow a total population of $\cN \cM_c \cM_s$ states.
The problem will be to see if the resulting algorithm 
reaches accurate results within the numerically
accessible values of $\cN, \cM_c, \cM_s$.

Some variants of the method should also be explored. In
particular we have not exploited the 
variational formulation in the computation of the probability $Q(\bh)$.  
One could also study 
in details the shape of the probability distributions
of local fields  in order to understand how they 
could parametrized in a simple but efficient way so that the free 
energy to be minimized 
does not involve a too large number of parameters.

We thank Giulio Biroli, Cristiana Carrus,  Enzo Marinari, R\'emi Monasson,
Riccardo Zecchina and Francesco Zuliani for useful discussions and comments.

\section*{Appendix I: Weighted sums of uncorrelated variables}
In this appendix we want to prove a useful little theorem, which applies to the 
computation of $F^{(1)}$ given in (\ref{fav1}) 
and $F^{(2)}$ in (\ref{fav2}).  

{\bf Theorem:} consider a set of $\cM (\gg 1)$ 
iid random free energies $f_\alpha$, ($\alpha \in \{ 1 \ldots M \} $)
 distributed with the exponential 
density \form{freedist} , and a set of 
$\cM$  positive numbers $a_\alpha$. Then, neglecting terms 
which go to zero when $ M$ 
goes to infinity, the following relation holds:
\be
\la \ln\(({\sum_\alpha a_\alpha \exp(-\beta f_\alpha) \over
\sum_\alpha  \exp(-\beta f_\alpha)} \)) \ra_{f}
\equiv 
\la\ln\((\sum_\alpha  w_{\alpha} a_\alpha\)) \ra_{f}=
{1 \over x} \ln \(({1 \over M} \sum_\alpha a_\alpha^x \)) \label{THEO}
\ee
where $\la.\ra_{f}$ denotes an average over the distribution of $f$.

{\bf Corollary 1:} In the same conditions as the theorem, for any set
of $M$ real numbers $b_\alpha$, one has:
\be
\la{\sum_\alpha a_\alpha b_\alpha \exp(-\beta f_\alpha) \over
\sum_\alpha a_\alpha \exp(-\beta f_\alpha)} \ra_{f}
=
{\sum_\alpha a_\alpha^x b_\alpha  \over
\sum_\alpha a_\alpha^x }
\ee

{\bf Corollary 2:} In the same conditions as the theorem, for any set
of $M$ real numbers $b_\alpha$, one has:
\be
\la{\sum_\alpha a_\alpha b_\alpha^2 \exp(-\beta f_\alpha) \over
\sum_\alpha a_\alpha \exp(-\beta f_\alpha)} \ra_{f}
-\la \(({\sum_\alpha a_\alpha b_\alpha \exp(-\beta f_\alpha) \over
\sum_\alpha a_\alpha \exp(-\beta f_\alpha)}²\))^2 \ra_{f}
=x \[[
{\sum_\alpha a_\alpha^x b_\alpha^2  \over
\sum_\alpha a_\alpha^x }- \(({\sum_\alpha a_\alpha^x b_\alpha  \over
\sum_\alpha a_\alpha^x }\))^2 \]]
\ee
 
{\bf Corollary 3:}
If the numbers $a_\alpha$ are $M$
 iid positive  random variables, such that the average of
$a^x$ exists, 
which are uncorrelated with the $f_\alpha$,
then one has:
\be
\la \ln\(({\sum_\alpha a_\alpha \exp(-\beta f_\alpha) \over
\sum_\alpha  \exp(-\beta f_\alpha)} \)) \ra_{f}
\equiv 
\la\ln\((\sum_\alpha  w_{\alpha} a_\alpha\)) \ra_{f}=
{1 \over x} \ln \(( \la  a_\alpha^x \ra_{a} \)) \label{THEO_coro}
\ee
where $\la.\ra_{a}$ denotes an average over the 
distribution of $a$.

{\bf Proof:}  we follow some of the techniques exposed in  \cite{MPV}.
We start from the identity
\be
\ln\((\sum_\alpha \exp(-\beta f_\alpha)  a_\alpha \))=\int \frac{dt}{t} 
\[[\exp(-t)-\exp\((-t\sum_\alpha\exp(-\beta f_\alpha)a_\alpha \))  \]] \ .
\ee
We choose to work with a regularised distribution of the $M$
iid random variables $f_\alpha$:
\be
P(f_\alpha)= \beta x \exp(\beta x (f_\alpha-f_c)) \ \theta(f_c-f) \ ,
\ee
where in the end we shall send $M \to \infty$, $f_c \to \infty$,
with $r= M \exp(-\beta f_c)$ fixed (the value of $r$ is irrelevant).
In this limit one has:

\be
\la \exp\((-t\exp(-\beta f_\alpha)a_\alpha \))\ra_f \simeq
1-(t a_\alpha)^x \exp(-\beta x  f_c) \Gamma(1-x) \ ,
\ee
from which one deduces:
\bea
 \la \ln\((\sum_\alpha \exp(-\beta f_\alpha)  a_\alpha) \)) \ra_f &=&
\int \frac{dt}{t} 
\[[\exp(-t)- \exp(-\Gamma(1-x) t^x \exp(-\beta x  f_c) \sum_\alpha a_\alpha^x \]]
 \\
&=&
{1 \over x} \ln \(( \Gamma(1-x) t^x M \exp(-\beta x  f_c) \sum_\alpha a_\alpha^x  \))
+{1-x \over x} C \ ,
\eea
where $C$ is Euler's constant. The quantity we need to compute involves 
subtracting the same expression with $a_\alpha$ substituted by one, which gives
the desired result:
\be
 \la \ln\((\sum_\alpha \exp(-\beta f_\alpha)  a_\alpha \))
- \ln\((\sum_\alpha \exp(-\beta f_\alpha) \)) \ra_f=
{1 \over x} \ln \(( {1 \over M} \sum_\alpha a_\alpha^x \))
\ee

The proof of corollary 1 is easily obtained by applying the theorem
to the set of numbers $a_\alpha \exp(\lambda b_\alpha)$,
and taking the derivative with respect to $\lambda$ at 
$\lambda=0$. Similar generalised formulas can be obtained
by taking higher order derivatives. The second derivative 
gives corollary 2. 
Corollary 3 is trivial.

{\bf Remark:}
We notice that in the two limits $x \to 0 $ and $x\to 1$ the equations can be simply 
understood:
\begin{itemize}
    \item
    In the limit $x=0$, in a typical realization of the random  free energies,
 only one weight $w$ is equal to one and all the others are zero. 
Averaging over the 
realizations of free energies amounts to spanning uniformly the set of indices
of this special non zero weight.
\item
    In the limit $x=1$ the number of relevant $w$ goes to infinity and each 
individual 
    contribution goes to zero. An infinite number of 
    term is present in the l.h.s. of \form{THEO} and the 
    r.h.s. of the \form{THEO} becomes 
$\ln \[[ (1/M) \sum_\alpha  a_\alpha \]]$, as it should.
\end{itemize}

\end{document}